\documentclass{article}

\usepackage{arxiv}
\usepackage{amsthm, amsmath, amsbsy, amssymb, booktabs, epsfig, epsf}
\usepackage{multicol, multirow, psfrag, booktabs, graphicx, subfig, rotating}
\usepackage[authoryear]{natbib}
\usepackage[colorlinks, citecolor = blue, urlcolor = blue]{hyperref}
\usepackage{tikz}

\newcommand{\PR}[1]{\ensuremath{\left[#1\right]}}
\newcommand{\PC}[1]{\ensuremath{\left(#1\right)}}
\newcommand{\chav}[1]{\ensuremath{\left\{#1\right\}}}
\newcommand{\bm}[1]{{\boldsymbol {#1}}}

\newcommand*\circled[1]{\tikz[baseline=(char.base)]{
            ~\node[shape=circle, draw, inner sep=0.75pt] (char) {#1};}}

\title{Alleviating Spatial Confounding in Spatial Frailty Models}

\author{
  Douglas  R. M. Azevedo \\
  Department of Statistics\\
  Universidade Federal de Minas Gerais\\
  Av. Presidente Ant\^{o}nio Carlos 6627 \\
  Pampulha Belo Horizonte, Minas Gerais \\
  31270-901, Brazil\\
  \texttt{douglasrm.azevedo@gmail.com} \\
  \And
  Marcos O. Prates \\
  Department of Statistics\\
  Universidade Federal de Minas Gerais\\
  Av. Presidente Ant\^{o}nio Carlos 6627 \\
  Pampulha Belo Horizonte, Minas Gerais \\
  31270-901, Brazil\\
  \texttt{marcosop@est.ufmg.br} \\
  \And
  Dipankar Bandyopadhyay \\
  Department of Biostatistics \\ 
  Virgina Commonwealth University \\
  830 E. Main Street, PO Box 980032 \\
  Richmond, VA \\
  \texttt{bandyop@vcuhealth.org} \\
}

\begin{document}
\maketitle

\begin{abstract}
Spatial confounding is how is called the confounding between fixed and spatial random effects. It has been widely studied and it gained attention in the past years in the spatial statistics literature, as it may generate unexpected results in modeling. The projection-based approach, also known as restricted models, appears as a good alternative to overcome the spatial confounding  in generalized linear mixed models. However, when the support of fixed effects is different from the spatial effect one, this approach can no longer be applied directly. In this work, we introduce a method to alleviate the spatial confounding for the spatial frailty models family. This class of models can incorporate spatially structured effects and it is usual to observe more than one sample unit per area which means that the support of fixed and spatial effects differs. In this case, we introduce a two folded projection-based approach projecting the design matrix to the dimension of the space and then projecting the random effect to the orthogonal space of the new design matrix. To provide fast inference in our analysis we employ the integrated nested Laplace approximation methodology. The method is illustrated with an application with lung and bronchus cancer in California - US that confirms that the methodology efficiency.
\end{abstract}

\section{Introduction} \label{sec:introduction} 

Frailty models is a useful and flexible class of models that allow the inclusion of latent effects related to an individual or a group in order to accommodate possible non-observed covariates. The most simple way to introduce a frailty term in a model is through an unstructured effect ensuring that the hazard is always positive for all sample units. However this approach does not handle with structured effects including the case of spatial data.

Papers as \citet{henderson2002modeling}, \citet{li2002modeling} and \citet{banerjee2003frailty} propose the use of frailty models to incorporate spatial structure into the latent effects. The latter used consolidated spatial models as the conditional autoregressive model (CAR) \citep{besag1974spatial} for areal data and the Gaussian model for georeferenced data \citep{cressie1992statistics}. 

Spatial models are widely studied in the literature and are important in practice to modeling spatially correlated data. In addition to the development of robust models, there are works focused on identifying (and solving) any limitations of such kind of approach. A possible limitation of spatial models is the spatial confounding \citep{reich2006effects, hughes2013dimension, hanks2015restricted, thaden2018structural, prates2018alleviating}. This problem resembles the multicollinearity in linear models that can distort the results and even lead to wrong conclusions. Spatial confounding occurs when the spatial effect brings similar information to that coming from the fixed effects. Thus, the point estimate of the regression coefficient related to the covariate becomes biased and leads to variance inflation.

Advances in technology, data storage and data quality allow the fitting of increasingly complex models aiming for better fit and more interpretability. To the best of our knowledge the spatial confounding in frailty models for areal data is an unexplored field which needs attention due to the importance of such models mainly for health sciences. The usual approaches for areal data cannot be directly applied for such kind of model as a result of the difference in the support of the spatial structure (areal level) and the fixed effects (sample unit level). Moreover in some applications there is a huge number of individuals in each area which leads to a computational challenge using this models' family. 

In this paper we provide a tool to alleviate the impact of spatial confounding in frailty models that works for a wide list of basal hazard functions and censoring schemes. The computational limitation is overcame using a trick that allow us to relief the spatial confounding issue using a reduction of the data dimension instead of working in the original one. We also propose a variance inflation factor (VIF) measure through the posterior sample of the spatial model (restricted and unrestricted). This measure allow us to validate the existence of spatial confounding and identify the covariates that are affect. As an application we have the lifetime of 72,612 individuals suffering from respiratory cancer (lung and bronchus) in California - US. This dataset is provided by SEER (surveillance, epidemiology, and end results program) \citep{seer}.

This paper is organized as follows. In section (\ref{sec:data_sources}) we present the dataset used in this work. In section (\ref{sec:methodology}) we have a review of spatial frailty models and spatial confounding. We show the proposed solution theory and the mathematical challenges of the restricted spatial frailty models in section (\ref{sec:restricted_spatial_frailty_models}). For model check and illustration, simulations and an application to the SEER respiratory cancer are performed in sections (\ref{sec:simulation}) and (\ref{sec:application}) respectively. Section (\ref{sec:final_remarks}) is reserved for conclusion and general discussion about the model and future work.

\section{Bronchus and lung cancer in California - US} \label{sec:data_sources} 

According to the Surveillance, Epidemiology, and End Results Program (SEER) \citet{seer}, lung and bronchus cancer reaches about $220.000$ men and women per year in the US. Just $20\%$ of them are expected to be alive five years after diagnosis. Therefore, it is disease with high mortality rates that must be investigated carefully.  

Our goal is to model the time until death by lung and bronchus cancer in California - US. To motivate our contributions we use individual-level as well as areal-level covariates to characterize individuals and regions in study. Therefore, two main data sources are in use in our analysis and are described below. 

\subsection{Datasets}

To add county-level information we are using the County Health Rankings \& Roadmaps (CHRR) \citep{chrr} which provides several important indices collected from different sources in the US. The CHRR is provided by the University of Wisconsin, Population Health Institute and is used to include inputs that describe characteristics of each California's county. The proportion of adults that smoke every day or most days between 2010 and 2016 for every California's county is accounted as an important risk factor for our analysis. Additional information about covariates as well as the dataset can be obtained in the CHRR website \href{https://www.countyhealthrankings.org/}{https://www.countyhealthrankings.org/}.

The main database in this work is provided by the SEER. The SEER program collects data on cancer cases from several locations and sources since 1973 in the United States. The data is provided by the National Cancer Institute (NHI) the American national leader in cancer research. The SEER datasets are not publicly available for free download. However, it can be requested following NIH/NCI SEER data access options via the link \href{https://seer.cancer.gov/data/options.html}{https://seer.cancer.gov/data/options.html}.

\subsection{Time until death by bronchus and lung cancer}

The data set provided by SEER has cases of lung and bronchus cancer for 72,612 individuals (between 2010 and 2016 after data cleaning) in the California state. In this data set, important covariates are present as gender and the stage of the disease for each individual. However, important covariates are missing as the case of tobacco consumption. To work around this problem, we are using the areal level covariate that indicates the percentage of adults that smoke every day or most days in each county. Because people can start in the program in different years (2010 - 2016), we picked the corresponding annual statistic in the CHRR dataset to be a proxy of how likely the person is a smoker.

Table~\ref{tab:survival_covariates} presents some summary information about the individual covariates used from the SEER program as well as the areal level covariate. Continuous covariates are represented by median (quantiles 25\% and 75\%) and categorical variables are represented by its observed proportion.

\begin{table}[hhh]
\centering
\caption{Summary statistics of SEER and CHRR covariates. For categorical variables the sample size and the percentage. For continuous variables median and quantiles 25\% and 75\%.}
\label{tab:survival_covariates}
\begin{tabular}{@{}ll@{}}
\toprule
Variable         & N = 72612         \\ \midrule
Time until event/censure & 10.0 [4.00; 25.0] \\
Status           &                   \\
~~~~ 0           & 31013 (42.7\%)    \\
~~~~ 1           & 41599 (57.3\%)    \\
Gender           &                   \\
~~~~ Female      & 34625 (47.7\%)    \\
~~~~ Male        & 37987 (52.3\%)    \\
Race             &                   \\
~~~~ Non-black   & 66723 (91.9\%)    \\
~~~~ Black       & 5889 (8.11\%)     \\
Cancer stage     &                   \\
~~~~ In situ     & 519 (0.71\%)      \\
~~~~ Localized   & 15870 (21.9\%)    \\
~~~~ Regional    & 16792 (23.1\%)    \\
~~~~ Distant     & 39431 (54.3\%)    \\
Age at diagnosis & 69.0 [62.0; 77.0] \\
\% Smokers       & 0.14 [0.12; 0.15] \\ \bottomrule
\end{tabular}
\end{table}
The time is measured in months being the median 10 months. The time variable was scaled (time/max$\{$time$\}$) in our models to avoid computational instabilities. 

For each individual, we assign a status of 1 if the individual died by lung or bronchus cancer and zero if the individual died by other causes or is still alive. Therefore, it can be seen as a right censoring scheme with about 43\% censured cases. Given the current dataset structure, it can be seen as a type I censoring scheme.

In our sample we have more cases of lung and bronchus cancer for men than for women as expected following the literature \citep{seer}. For simplicity the race variable was coded as black and non-black for each individual. ``In situ'' refers to abnormal cells that are present in the lung or bronchus but have not spread to nearby. ``Located'' corresponds to the stage where the cancer is limited to where it started and has not spread yet. In the ``regional'' phase, cancer has spread to nearby lymph nodes or organs. The last and more severe phase is the ``distant'' stage. In the ``distant'' stage, cancer has spread to distant parts of the body. Therefore, we expect an increase in the risk of death following this logical order. Finally, the median age is 69, which corresponds to an elderly population. 

\section{Methodology} \label{sec:methodology}

\subsection{Spatial models}\label{sec:spatial_models}

Modeling the sources of variation is important for countless fields and can help researchers identifying spatial patterns and make decisions. In many cases, the data is spatially structured which makes the models capable of identifying the influence of these structures indispensable.

In Statistics the most common types of spatial data are those in which observations are collected in a continuous space (geostatistical data) or when they represent a region in space (areal data). In the areal data context, it is commonly incorporated the neighborhood structure information into modeling to capture the spatial heterogeneity. This is done to capture the spatial behavior of a possible unobserved or latent covariate. 

There are several approaches for modeling spatially structured data in the areal context as the cases of conditional autoregressive (CAR) \citep{besag1974spatial, banerjee2014hierarchical}, simultaneous autoregressive (SAR) \citep{whittle1954stationary, ord1975estimation}, Leroux \citep{leroux1999estimation}, mixture neighborhood structure \citep{rodrigues2012bayesian} and, directed acyclic graph autoregressive (DAGAR) \citep{datta2018spatial}. 

In applied sciences, the most common approach for areal data is the Intrinsic CAR (ICAR) \citep{besag:york:mollie} model due to the model simplicity. Although any of the mentioned methodologies are valid and can be easily explored in our framework, in this work we adopt the ICAR model.

Let $Y_1, \ldots, Y_n$, random variables observed in $n$ areas and take $\mathbf{\Theta} = \PC{\theta_1, \ldots, \theta_n}^T$, random effects with zero mean, related to each location. The traditional ICAR defined by CAR($1$, $\frac{1}{\sigma^2}$) model defines the conditional distributions as
\begin{equation*} \label{eq:car_improper}
 \PC{\theta_i|\mathbf{\theta_{-i}}} \sim N\PC{\sum_{j \sim i}\frac{\theta_j}{w_{i+}}, \frac{\sigma^2}{w_{i+}}},
\end{equation*}
where $\mathbf{\theta_{-i}}$ represents the vector $\mathbf{\Theta}$ except the element $\theta_i$, and $j\sim i$ indicates that the areas $i$ and $j$ are neighbors. $\mathbf{W}$ is the adjacency matrix, where $w_{ij} = 1$ if and only if $i$ and $j$ are neighbors and $0$ elsewhere. The notation $w_{i+}$ indicates the sum of the elements of the $i_{th}$ line of $\mathbf{W}$, and therefore represents the number of neighbors in the $i_{th}$ region. 

If we change the expected mean to $\displaystyle \sum_{j\sim i}\rho\frac{\theta_j}{w_{i+}}$ this model is called CAR($\rho$, $\frac{1}{\sigma^2}$) and $\rho$ is the spatial dependence parameter responsible to indicates the dependence intensity between regions \citep{banerjee2014hierarchical}.

The CAR dependence matrix structure is given by $\bm{Q} = \tau(\bm{D_w} - \rho\bm{W})$ where $\tau = \frac{1}{\sigma^2}$ and $\bm{D_w}$ is a diagonal matrix with entries $w_{i+}$ representing the number of neighbors of each region. The ICAR prior model is obtained by setting 
$\rho = 1$.


\subsection{Spatial confounding}\label{sec:spatial_confounding}

A current limitation in spatial statistics is the so-called spatial confounding. This problem resembles what occurs in linear models when two or more covariates bring the same information about the response variable. In linear models, this is called multicollinearity and inflates the variance of the regression coefficient estimators. This inflation, in some cases, changes the model interpretation leading the researcher, occasionally, to get to wrong conclusions about the necessity and importance of the covariate in the model. 

An extension of the multicollinearity problem to the spatial context is the so called spatial confounding and it occurs when the spatial effect brings similar information to one or a linear combination of the covariates in the model. Differently from the multicollinearity, spatial confounding determines an inflation in the variance of the regression estimators and also a bias in the point estimate, possibly changing conclusions drastically \citep{reich2006effects}.

Recently, several works as \citep{reich2006effects, hughes2013dimension, hanks2015restricted, hefley2017bayesian, guan2018computationally, thaden2018structural, prates2018alleviating} approached the spatial confounding problem either for generalized linear mixed models areal or geostatistical data. \cite{reich2006effects} mathematically formulated the problem of spatial confounding for spatial linear regression models. The authors employed the ICAR prior in the following modelling:
\begin{align} \label{eq:reich_model}
  & \bm{Y}|\bm{\beta}, \bm{\psi}, \tau_{\epsilon} \sim \text{Normal}(\bm{X\beta} + \bm{\psi}, \tau_{\epsilon}\bm{I}_n), \\
  & \bm{\psi}|\tau_{\psi} \sim \text{Normal}(0, \tau_{\psi}\bm{Q}), \nonumber
\end{align}
\noindent where $\bm{\psi}$ is the ICAR spatial effect, $\bm{Q}$ is the ICAR precision matrix aforementioned, $\tau_{\epsilon}$ and $\tau_{\psi}$ are precision parameters related to the Gaussian observations and the ICAR, respectively. Note that in our parametrization the Gaussian distribution is being represented by its precision matrix rather than its covariance matrix.

In the case of spatial linear regression, it is possible to analytically calculate the expected mean and variance integrating out the latent effect as:
\begin{align} \label{eq:reich_model_mean_variance}
  & E(\bm{\beta}|\tau_{\epsilon}, \tau_{\psi}, \bm{y}) = \displaystyle (\bm{X^TX})^{-1}\bm{X}^T(\bm{y} - \hat{\bm{\psi}}) = \bm{\beta}_{ols} - (\bm{X^TX})^{-1}\bm{X}^T\hat{\bm{\psi}}, \\
  & Var^{-1}(\bm{\beta}|\tau_{\epsilon}, \tau_{\psi}, \bm{y}) = \displaystyle \tau_{\epsilon}(\bm{X^TX}) - \bm{X}^T Var(\bm{\psi}|\bm{\beta}, \tau_{\epsilon}, \tau_{\psi}, \bm{y})\bm{X}, \nonumber 
\end{align}
where $\bm{\beta}_{ols} = (\bm{X^TX})^{-1}\bm{X}^T\bm{y}$ and $\hat{\bm{\psi}} = E(\bm{\psi}|\tau_{\epsilon}, \tau_{\psi}, \bm{y})$.

Carefully investigating Equation~\eqref{eq:reich_model_mean_variance} it is possible to conclude that the predicted $\bm{y}$ value is given by $\bm{X}(\bm{X^TX})^{-1}\bm{X}^T(\bm{y} - \bm{\hat{\psi}}) = \bm{P}(\bm{y}- \hat{\bm{\psi}}) = \bm{Py} - \bm{P}\hat{\bm{\psi}}$, a projection of $\bm{y}$ onto the space of $\bm{X}$ minus a projection of the latent effect $\hat{\bm{\psi}}$ onto the space of $\bm{X}$.

When the spatial confounding is present it indicates the existence of duplicated information in the model. One way to alleviate this problem is by using a projection-based approach. That is done by decomposing the spatial effect into the projection onto the space of the covariates and the projection onto the orthogonal space of the covariates as in Equation~\eqref{eq:projection_based}.
\begin{align} \label{eq:projection_based}
  & \bm{Y}|\bm{\beta}, \bm{\psi}, \tau_{\epsilon} \sim \text{Normal}(\bm{X\beta} + \bm{P\psi} + \bm{P^{\bot}\psi}, \tau_{\epsilon}\bm{I_n}), \\
  & \bm{\psi}|\tau_{\psi}, \sim \text{Normal}(0, \tau_{\psi}\bm{Q}), \nonumber
\end{align}
in which, $\bm{P^{\bot}} = (\bm{I} - \bm{P})$ is the projection matrix onto orthogonal space of $\bm{X}$.

Therefore, in Equation~\eqref{eq:projection_based} $\bm{P\psi}$ corresponds to the information of $\bm{\psi}$ on the space of $\bm{X}$ which in other words represents the duplicated information. Thus, one way to alleviate the spatial confounding is by removing $\bm{P\psi}$ from the model. 

This approach is known as restricted spatial regression (RSR) and is also applicable for generalized linear mixed models (GLMMs). However, for GLMMs, it is not possible to analytically evaluate the impacts of the latent effect on the coefficients estimates as the analytical solution of the involved integrals are not available. Hereafter, we will refer to this approach as the RHZ model.

\cite{hanks2015restricted} focused their effort into geostatistical data instead of areal data as the previous work. Also, the authors reported several simulation studies about inference under model misspecification. For geostatistical data, one must assume that the correlation structure may be a function of the distance between points in the space and some parameters might govern the spatial relationship between areas. One of the most common covariance structure for continuous spatial correlation is the Mat\'{e}rn structure \citep{cressie1992statistics}. In this model, each $\bm{\Sigma}$ (covariance matrix) entry of is defined as
$$\Sigma_{ij} = \sigma^2\frac{1}{\Gamma(\nu)2^{\nu-1}}\PC{\sqrt{2\nu}\frac{d_{ij}}{\phi}}^{\nu}K_{\nu}\PC{\sqrt{2\nu}\frac{d_{ij}}{\phi}}$$
where $d_{ij}$ is the Euclidean distance between the $i_{th}$ and the $j_{th}$ observations, $\sigma^2$ is the partial sill parameter, $\nu$ is a smoothness parameter, $\phi$ is the range parameter, and $K_{\nu}$ is the modified Bessel function of the second kind.

In the RHZ model, $\bm{\Sigma} = \bm{Q}^{-1}$ is fixed, then it is possible to compute $\bm{Q}$ just once. However, in the continuous case, the matrix $\bm{\Sigma}$ may vary given the parameters $\{\sigma^2, \nu, \phi\}$. Thus, an approach similar to the previously mentioned may not be feasible because each step of MCMC (or the step in a numerical optimization routine) would require the matrix $\bm{Q} = \bm{\Sigma}^{-1}$ to evaluate the likelihood. To obtain an efficient algorithm \cite{hanks2015restricted} suggest the use of the conditioning by kriging technique \citep{rue2005gaussian}. The idea is to sample from the unrestricted model $\bm{\psi^{\ast}} \sim \text{Normal}(\bm{0}, \bm{\Sigma})$ and then to have a sample, under the restriction $\bm{P\psi} = \bm{0}$, take $\bm{\psi} = \bm{\psi}^{\ast} - \bm{\Sigma X}(\bm{X\Sigma X})^{-1}\bm{X}^T\bm{\psi}^{\ast}$.

Through simulations, the authors showed that using the conditioning by kriging technique the inference for fixed effects is more appropriate by comparing the Type-S error \citep{gelman2000diagnostic}. A Type-S error occurs when the regression parameters are equal to zero and the $95\%$ posterior credibility interval does not contain the zero value.

Another important contribution of this work is the possibility to get a sample from both models (restricted and unrestricted) concurrently, along the MCMC. This is possible because there is an equivalence between the RHZ model and the unrestricted model:
\begin{align*} 
  E(Y_i|\bm{\beta}) &= \bm{X\beta}_{rsr} + \bm{\psi}_{rsr} \\
               &= \bm{X\beta}_{rsr} + \bm{(I-P)\psi}_{sr} \\
               &= \bm{X\beta}_{rsr} + \bm{\psi}_{sr} - \bm{P\psi}_{sr} \\
               &= \bm{X\beta}_{rsr} + \bm{\psi}_{sr} - \bm{X(X^TX)^{-1}X^T}\psi_{sr} \\
               &= \bm{X(\beta}_{rsr} - (\bm{X}^T\bm{X})^{-1}\bm{X}^T\bm{\psi}_{sr}) + \bm{\psi}_{sr} \\
                &= \bm{X\beta}_{sr} + \bm{\psi}_{sr},
\end{align*}
where $\bm{\beta}_{rsr}$ refers to the coefficients of the restricted model proposed by \cite{reich2006effects}, $\beta_{sr}$ corresponds to the unrestricted model, $\bm{\psi}_{rsr}$ are the latent effects of the restricted model and $\bm{\psi}_{sr}$ the latent effects of the unrestricted model.

\subsubsection{Measures of spatial confounding} \label{subsec:svif}

A common confounding measure for spatial models is the spatial VIF (SVIF) as proposed by \cite{reich2006effects}. This measure is equivalent, for each $\beta_j$, to the ratio between the variance of $\beta_j$ for the spatial model and the variance for the model without spatial component as in Equation~\eqref{eq:VIF_regression}. This measure reflects the increment in the variance after adding the spatial component. 

For linear regression models, it is possible to calculate the exact value of these two quantities (Equation~\eqref{eq:reich_model_mean_variance}). \cite{reich2006effects} also note that this measure depends only on $r = \frac{\tau_{\psi}}{\tau_{\epsilon}}$ where $\tau_{\epsilon}$ is the precision of the Gaussian response, being the scale of the latent effect important in such kind of study \citep{paciorek2010importance}.

The SVIF is defined as:
\begin{equation}\label{eq:VIF_regression}
    SVIF(\beta_j|r, \tau_{e}) = \frac{Var(\beta_j)_{slr}}{Var(\beta_j)_{lr}},
\end{equation}
where $Var(\beta_j)_{slr}$ is the variance of the coefficient for the spatial linear regression and $Var(\beta_j)_{lr}$ is the variance for the linear regression.

Under a GLMM it is not possible to derive a closed form for $Var(\beta_j)$, then the solution is to approximate these quantities by the Fisher information. The same occurs for frailty models and in this case one may use the Fisher information or the empirical variance obtained in a posterior sample (in Bayesian frameworks). Thus one can calculate the SVIF as
\begin{equation}\label{eq:VIF_frailty}
    SVIF(\beta_j) = \frac{Var(\beta_j)_{sm}}{Var(\beta_{j})_{m}},
\end{equation}
where $Var(\beta_j)_{sm}$ is the sample variance of the coefficient for the spatial model and $Var(\beta_j)_{m}$ is the sample variance for the model without the spatial component.

With the $SVIF(\beta_j)$, one can compare two models and investigate if the variance is inflated after the latent effect inclusion. However, we can also evaluate the effectiveness of the restricted model. An equivalent way to measure the variance's impact is by using the variance retraction factor defined as:
\begin{equation}\label{eq:VRF_frailty}
    SVRF(\beta_j) = \frac{Var(\beta_j)_{u} - Var(\beta_j)_{r}}{Var(\beta_j)_{u}},
\end{equation}
where $Var(\beta_j)_{u}$ is the variance of the coefficient for the unrestricted model and $Var(\beta_j)_{r}$ is the variance for the restricted model.

This measure is zero if $Var(\beta_j)_u = Var(\beta_j)_r$, greater than zero if $Var(\beta_j)_u > Var(\beta_j)_r$ and less than zero otherwise. A $SVRF(\beta_j) = 0.4$ can be interpreted as a $40\%$ retraction in the coefficient variance under the restricted model in comparison with the unrestricted one.

\subsection{Spatial frailty model}

\subsubsection{Survival models}

Survival models are an important tool in several branches of science mainly in health data analysis. In general, the researcher is interested in using survival models to answer questions about phenomena that can be measured in units of time. Since the response is observed in units of time, it is assumed for it, distributions with support in the positive real numbers. The most commonly employed models make use of simple probability distributions such as exponential, gamma, lognormal or Weibull. More complex models rely on less conventional distributions such as the Birnbaum-Saunders \citep{birnbaum1969new} or semi-parametric approaches as in the case of the piecewise exponential model \citep{friedman1982piecewise}.

A great differential of survival models is that in practice the phenomenon of interest is not always observed. To deal with this situation, without loss of information, it is necessary to take advantage of a censoring scheme. There are several censoring schemes in the literature being the right, left and interval censoring schemes the most famous ones. 
Besides that, the censoring may happen according to some mechanism being the most famous ones the type I and type II censoring mechanisms. The former happens when the study has a fixed endpoint and then people who have not yet experienced the event are censured. The latter occurs when a number of events, defined a priori, is achieved. Thus, all other individuals are censored. For a review of survival models see \cite{hosmer2008applied}. 

In several cases, as the cited ones, the likelihood fits in Equation~\eqref{eq:likelihood_frailty}.
\begin{align} \label{eq:likelihood_frailty}
    \mathcal{L}(\bm{\theta}; t) = &\prod_{d\in \bm{D}}f_{\bm{\theta}}(t_d) \\
                   &\prod_{r\in \bm{R}}S_{\bm{\theta}}(t_r) \prod_{l\in \bm{L}}\PC{1-S_{\bm{\theta}}(t_l)} \nonumber \\
	               &\prod_{k\in \bm{K}}\PC{S_{\bm{\theta}}(t_{k1}) - S_{\bm{\theta}}(t_{k2})} \nonumber,
\end{align}
\noindent where $\bm{D}$ is the set of observed failure/event times, $\bm{R}$ is the set of right-censored sample units, $\bm{L}$ is the set of left-censored sample units and $\bm{K}$ is the set of interval-censored sample units. The time until failure or the censoring time is denoted by $t$ for left and right censoring schemes. For interval censoring, two times are provided and then we denote the lower bound of this interval as $t_{k1}$ and the upper bound as $t_{k2}$ for a sample unit $k$. The distribution assumed for the phenomenon of interest is represented by $f_{\bm{\theta}}(\bm{t})$ (in parametric models) and $S_{\bm{\theta}}(\bm{t})$ is the survival function. 

As common choices for $f_{\bm{\theta}}(\bm{t})$ we can cite exponential, gamma, lognormal and Weibull distributions. The function choice leads to different forms of the $h_{\bm{\theta}}(\bm{t})$ called hazard function. It measures the instantaneous risk of occurrence of an event. $S_{\bm{\theta}}(\bm{t})$ is the survival function which indicates the probability of occurrence of the event at a time $T > t$. Any other parameters of $f_{\bm{\theta}}(t)$ are represented by the vector $\bm{\theta}$.

The functions $f_{\bm{\theta}}(t)$, $S_{\bm{\theta}}(t)$ and $h_{\bm{\theta}}(t)$ are linked through the following identities:
\begin{align} \label{eq:survival_identities}
    & S_{\bm{\theta}}(t) = \Pr(T > t) = \int _{t}^{{\infty }}f_{\bm{\theta}}(u)\,du = 1-F_{\bm{\theta}}(t), \nonumber \\
    & h_{\bm{\theta}}(t)=\lim _{dt\rightarrow 0}{\frac {\Pr(t\leq T<t+dt)}{dt}\times\frac{1}{S_{\bm{\theta}}(t)}}={\frac {f_{\bm{\theta}}(t)}{S_{\bm{\theta}}(t)}}=-{\frac {S_{\bm{\theta}}'(t)}{S_{\bm{\theta}}(t)}}, \\
    & f_{\bm{\theta}}(t)=h_{\bm{\theta}}(t)\times S_{\bm{\theta}}(t). \nonumber
\end{align}
The survival function, $S_{\bm{\theta}}(t)$, has the property that $S_{\bm{\theta}}(0) = 1$ and $S_{\bm{\theta}}(\infty) = 0$. However, in some cases, it is possible to observe that some individuals will never experience the event of interest because they may be not exposed to the phenomenon anymore. In those cases one can use a cure fraction model \citep{boag1949maximum} where there is a proportion of the individuals which will never experience the event of interest. The simpler way to introduce a cure fraction in the modeling is by a mixture model. In this case, the survival function is a mixture of a proper survival function and a point mass at a constant $c$, called cure fraction as above
$$S_{\bm{\theta}}(t) = c + (1-c)S^{\ast}_{\bm{\theta}}(t).$$
Other approaches for the cure fraction can be found in the literature as, for example, \citet{tsodikov2003estimating, lambert2007modeling, scudilio2019defective}.

In survival analysis, the interest is to model the hazard function to understand factors that impact the risk of an event. Therefore, covariates may be included into the model to measure their impact. Several parametric models are described in the literature as the cases in Table \ref{tab:survival_functions}. 
\begin{table}[h]
\caption{Distributions and its respective set of parameters ($\theta$), probability density function ($f_{\theta}$), survival function ($S_{\theta}$) and hazard function ($h_{\theta}$). The term $\Phi(t)$ represents the cumulative distribution function of a standard Normal distribution.}
\label{tab:survival_functions}
\centering
\begin{tabular}{@{}lllll@{}}
\toprule
Distribution & $\theta$               & $f_{\theta}(t)$             & $S_{\theta}(t)$ & $h_{\theta}(t)$  \\ \midrule
Exponential  & $\{\lambda\}$          & $\lambda\exp\{-\lambda t\}$ & $\exp\{\lambda t\}$ & $\lambda$ \\
Lognormal    & $\{\mu, \sigma\}$      & $\frac {1}{(2\pi)^{1/2}\sigma t}\exp\chav{-\frac{1}{2}\PC{\frac{log(t) - \mu}{\sigma^2}}^2}$     & $1-\Phi\PC{\frac{\log{t} - \mu}{\sigma}}$ & $\frac{f_{\theta}(t)}{S_{\theta}(t)}$ \\
Gamma        & $\{\alpha, \lambda\}$  & $\frac {\lambda ^{\alpha }}{\Gamma (\alpha )}t^{\alpha -1}e^{-\lambda t}$ & 1 - $\frac {1}{\Gamma (\alpha )}\int_0^{\lambda t} u^{\alpha -1}e^{-u}du$ & $\frac{f_{\theta}(t)}{S_{\theta}(t)}$  \\
Weibull      & $\{\alpha, \lambda\}$  & $\alpha\lambda t^{\alpha - 1}\exp\{\lambda t^{\alpha}\}$ & $\exp\{-\lambda t^{\alpha}\}$ & $\alpha\lambda t^{\alpha-1}$ \\ \bottomrule
\end{tabular}
\end{table}

A well studied method to include covariates in the modeling is the Cox proportional hazards model \citep{cox1972regression}. Its idea is to insert the covariates on the hazard function in a multiplicative way ensuring that the hazard is never negative. This model assumes proportional hazards meaning that the hazard ratio for two individuals is constant over time. Next equation shows the hazard function under the Cox proportional hazard model:
\begin{equation}
    h_{\bm{\theta}}(t_i|\bm{X}_i) = h^{\ast}_{\bm{\theta}}(t_i)\exp\{\bm{X}_i\bm{\beta}\},
\end{equation}
\noindent where $h^{\ast}_{\bm{\theta}}(t_i)$ is called baseline hazard function.

However, one can use the partial likelihood technique which makes the baseline hazard specification unnecessary \citep{cox1972regression}. Another alternative, is to create a fully parametric proportional hazard model by replacing $h^{\ast}_{\bm{\theta}}(.)$ by a parametric baseline hazard function \citep{lawless2011statistical}.

In many cases, the introduction of covariates is not enough for an appropriate fit. This is explained by the fact that often, important covariates are not observed or they are impossible to measure. In this case, one can introduce a latent effect giving rise to a frailty model.

Similarly to GLM models, one can introduce latent effects to take the non-observed covariates and/or clusters effects into consideration. This model family is known as frailty models \citep{wienke2010frailty}. In general, the easiest way to introduce these effects is in a multiplicative way. Because the hazard is a positive quantity it is necessary to guarantee that the multiplicative effect will ensure that the hazard is still positive. One way is to assume that $\bm{\gamma}$, the frailty term, is drawn from a positive probability distribution.
\begin{equation*} \label{eq:funcao_risco_fragildiade}
    h_{\bm{\theta}}(t_{ij}|\bm{X}_{ij}) = h^{\ast}_{\bm{\theta}}(t_{ij})\gamma_j\exp\{\bm{X}_{ij}\bm{\beta}\},
\end{equation*}
\noindent where $\gamma_j$ is called frailty (related to cluster $j$) and a common choice for its distribution is the gamma distribution, given rise to the gamma frailty model. However, under this distribution, it is difficult to insert dependence structures between clusters and then they are, in general, considered independent.

The main idea in spatial statistics is to insert dependence between geographically close locations. Using the gamma frailty model, the inclusion of spatial structure is not trivial. Several approaches in the literature try to deal with spatially structured effects in frailty models as \cite{henderson2002modeling}, \cite{li2002modeling} and \cite{banerjee2003frailty}. The latter proposed a frailty model that allows the insertion of already known structures of spatial models. In this case, the spatial effect enters the model in an additive way, but within the exponential term, which gives rise to the model presented in Equation~\eqref{eq:spatial_frailty_model}. Take $j = 1, \ldots, n_i $, indices of $n_i$ sample units observed in the location $i$ for $i = 1, \ldots, n $, $n$ locations. The hazard function of the spatial frailty model is given by:
\begin{align} \label{eq:spatial_frailty_model}
  h_{\bm{\theta}}(t_{ij}|\bm{X}_{ij}) & = h_{\bm{\theta}}^{\ast}(t_{ij}) \times \gamma_i \times  e^{\bm{X}_{ij} \bm{\beta}}\nonumber \\
  & = h_{\bm{\theta}}^{\ast}(t_{ij}) \times e^{\bm{X}_{ij}\bm{\beta} + log(\gamma_i)} \\
  & = h_{\bm{\theta}}^{\ast}(t_{ij}) \times e^{\bm{X}_{ij}\bm{\beta} + \psi_i} \nonumber,
\end{align}
\noindent where $\psi_i$ is Gaussian and consequently the vector $\bm{\psi}$ is a multivariate normal distribution. This setting is convenient since several spatial models use the multivariate normal distribution as in the case of this work.

\section{Restricted Spatial Frailty models} \label{sec:restricted_spatial_frailty_models}

\subsection{Method} \label{chap:survival}

The likelihood of the spatial frailty model depends on the hazard function which is related to the baseline hazard function, covariates, and latent effects. As a consequence, this likelihood can be written according to Equation~\eqref{eq:spatial_frailty_likelihood}. Let $h_{\bm{\theta}}^{0}$ and $H_{\bm{\theta}}^{0}$ be the baseline hazard function and the cumulative baseline hazard function, respectively. Let $\bm{\psi} = \PR{\psi_1, \ldots, \psi_n}^T$ be a vector of latent effects related to each location. Define $\bm{\epsilon}$ as a vector with entries $\epsilon_{ij}$ $i = 1, \ldots, n$; $j = 1, \ldots, n_i$ where $\epsilon_{ij}$ is an unstructured latent effect related to the sample unit $j$ at location $i$. The likelihood is given by
\begin{small}
\begin{align} \label{eq:spatial_frailty_likelihood}
& \mathcal{L}(t, \bm{\theta}, \bm{\beta}, \bm{\psi}, \bm{\epsilon}) = \prod_{i=1}^{n}\prod_{j=1}^{n_i} \Big[\PR{h_{\bm{\theta}}(t_{ij})S_{\bm{\theta}}(t_{ij})}^{\Delta_D}S_{\bm{\theta}}(t_{ij})^{\Delta_R}\PC{1-S_{\bm{\theta}}(t_{ij})}^{\Delta_L}\PC{S_{\bm{\theta}}(t_{ij1}) - S_{\bm{\theta}}(t_{ij2})}^{\Delta_K} \Big ], \nonumber \\
& h_{\theta}(t_{ij}) = h_{\theta}^0(t_{ij})\exp\chav{\bm{X}_{ij}\bm{\beta} + \psi_i + \epsilon_{ij}}, \\
  & S_{\theta}(t_{ij}) = \exp\chav{-H_{\theta}^0(t_{ij})\exp\chav{\bm{X}_{ij}\bm{\beta} + \psi_i + \epsilon_{ij}}}, \nonumber 
\end{align}
\end{small}

\vspace{-0.2cm}

\noindent and $\bm{\Delta} = \chav{\Delta_D, \Delta_R, \Delta_L, \Delta_K}$ are indicator functions of events, right-censored sample units, left-censored sample units and interval-censored sample units, respectively, and represent, for each individual, which term will contribute in the likelihood.

In this model there is more sample units than locations which implies in different supports for $\bm{X}_{N\times p}$ and $\bm{\psi}_{n\times 1}$, where $N = \sum_{i = 1}^n n_i$. As mentioned by \cite{hanks2015restricted}, the projection-based approach is intuitive when the support of the observations is identical to spatial support, but we might be careful when this is not true as in the case of spatial frailty models. That said, in the conventional projection-based approach, the projection matrix is given by $\bm{P}_{(N\times N)} = \bm{X}(\bm{X}^T\bm{X})^{-1}\bm{X}^T$ and, therefore, it is not possible to make the projection of $\bm{\psi}_{n\times 1}$ onto the orthogonal space of $\bm{X}_{N\times p}$ directly. The simpler solution is to create a new vector of the same length as $\bm{X}$ by repeating the spatial effects according to the areas where $\bm{X}$ were collected. Define $\bm{\Psi} = [\psi_1\times \bm{1}_{n_1}, \ldots, \psi_n\times \bm{1}_{n_n}]^T$ where $\bm{1}_{m}$ is a length $m$ row vector of ones. For notation simplification we are considering that the $\bm{X}$ matrix is sorted by regions althoug it is not necessary. Thus, $\bm{\Psi}$ is represented in Equation~\eqref{eq:big_spatial_effect}
\begin{equation} \label{eq:big_spatial_effect}
    \bm{\psi}_{n\times 1} =
    \begin{bmatrix}
        \psi_1     \\
        \psi_2     \\
        \vdots  \\
        \psi_n
    \end{bmatrix};
    ~~~~
    \bm{\Psi}_{N\times 1} =
    \begin{array}{c@{\!\!\!}l}
        \left[ 
            \begin{array}[c]{ccccc}
                \psi_1     \\
                \vdots  \\
                \psi_1     \\
                \vdots  \\
                \psi_n     \\
                \vdots  \\
                \psi_n
            \end{array}  
        \right]
        &
        \begin{array}[c]{@{}l@{\,}l}
            \left. 
                \begin{array}{c} 
                    \vphantom{0}      \\
                    \vphantom{\vdots} \\ 
                    \vphantom{0} 
                \end{array} 
            \right\} 
            & 
            \text{$n_1$ times} \\
            \vphantom{\vdots}  \\
            \left. 
                \begin{array}{c} 
                    \vphantom{0}      \\ 
                    \vphantom{\vdots} \\ 
                    \vphantom{0}  
                \end{array} 
            \right\} 
            & 
            \text{$n_n$ times}
        \end{array}
    \end{array}.
\end{equation}
Then we can rewrite the hazard function in terms of the new vector in matrix format.
\begin{equation} \label{eq:unrestricted_spatial_frailty}
  h_{\bm{\theta}}(t) = h_{\bm{\theta}}^0(t)\exp\chav{\bm{X}\bm{\beta} + \bm{\Psi} + \bm{\epsilon}}.
\end{equation}
Given this configuration, we can apply a projection-based approach and decompose the vector $\bm{\Psi}$ into $\bm{P\Psi}$ and $\bm{P^{\bot}\Psi}$ where $\bm{P}^{\bot} = (\bm{I} - \bm{P})$
\begin{equation} \label{eq:decomposed_frailty_model}
  h_{\bm{\theta}}(t) = h_{\bm{\theta}}^0(t)\exp\chav{\bm{X\beta} + \bm{P\Psi} + \bm{P}^{\bot}\bm{\Psi} + \bm{\epsilon}}.
\end{equation}

The duplicated information in Equation~\eqref{eq:decomposed_frailty_model} is the vector $\bm{P\Psi}$ and may promote the bias and variance inflation. To alleviate it, a convenient solution is to remove this quantity giving rise to the following model
\begin{equation}
  h_{\bm{\theta}}(t) = h_{\bm{\theta}}^0(t)\exp\chav{\bm{X\beta}_{rsf} + \bm{P}^{\bot}\bm{\Psi} + \bm{\epsilon}},
\end{equation}
\noindent as $\bm{P} = \bm{X}(\bm{X}^T\bm{X})^{-1}\bm{X}^T$, we expect that this information will be incorporated by the coefficients as $\bm{\beta}_{rsf} = \bm{\beta} + (\bm{X}^T\bm{X})^{-1}\bm{X}^T\bm{\Psi}$ where ``rsf'' means ``restricted spatial frailty''. However, this solution implies in a limitation to the model. The spatial effect ($\bm{P}^{\bot}\bm{\Psi}$), free of spatial confounding, is a $N\times 1$ vector which does not have a meaning as we have just $n$ locations. Therefore, we propose a summarization of this information creating two vectors. The first vector contains the means by regions of $\bm{P}^{\bot}\bm{\Psi}$ and the second one contains the deviations from these means.

Define $\bm{\psi}_{rsf} = \PR{\psi_{rsf_{1}}, \ldots, \psi_{rsf_{n}}}^T$ the vector containing the $n$ means of $\bm{P^{\bot}\Psi}$, one for each region $i = 1, \ldots, n$, and $\tilde{\bm{\psi}} = \PR{\tilde{\psi}_{11}, \ldots, \tilde{\psi}_{nn_n}}^T$, where $\tilde{\psi}_{ij}$ represents the individual distance of each element of $\bm{P^{\bot}\Psi}$ to its respective mean $\psi_{rsf_{i}}$. In this case, $\bm{\Psi}_{rsf} = \PR{\psi_{rsf_1}\times \bm{1}_{n_1}, \ldots, \psi_{rsf_n}\times \bm{1}_{n_n}}^T$ is a vector of remaining mean effects of each location and $\tilde{\bm{\psi}}$ represents, for each sample unit, an individual distance from the mean as in Equation~\eqref{eq:final_spatial_effect}. In this case, both $\bm{\Psi}_{rsf}$ and $\tilde{\bm{\psi}}$ are $N\times 1$ vectors

\begin{small}
\begin{equation} \label{eq:final_spatial_effect}
    \bm{\Psi}_{rsf} =
    \begin{array}{c@{\!\!\!}l}
        \left[ 
            \begin{array}[c]{ccccc}
                \psi_{rsf_1}     \\
                \vdots  \\
                \psi_{rsf_1}     \\
                \vdots  \\
                \psi_{rsf_n}     \\
                \vdots  \\
                \psi_{rsf_n}
            \end{array}  
        \right]
        &
        \begin{array}[c]{@{}l@{\,}l}
            \left. 
                \begin{array}{c} 
                    \vphantom{0}      \\
                    \vphantom{\vdots} \\ 
                    \vphantom{0} 
                \end{array} 
            \right\} 
            & 
            \text{$n_1$ times} \\
            \vphantom{\vdots}  \\
            \left. 
                \begin{array}{c} 
                    \vphantom{0}      \\ 
                    \vphantom{\vdots} \\ 
                    \vphantom{0}  
                \end{array} 
            \right\} 
            & 
            \text{$n_n$ times}
        \end{array}
    \end{array}; 
    ~~~~
    \tilde{\bm{\psi}} =
    \begin{bmatrix}
        \tilde{\psi}_{11}  \\
        \vdots        \\
        \tilde{\psi}_{11}  \\
        \vdots        \\
        \tilde{\psi}_{nn_n}
    \end{bmatrix},
\end{equation}
\end{small}
\noindent and then we can rewrite the model as in Equation~\eqref{eq:frailty_decomposition}
\begin{equation} \label{eq:frailty_decomposition}
  h_{\bm{\theta}}(t) = h_{\bm{\theta}}^0(t)\exp\chav{\bm{X\beta}_{rsf} + \bm{\Psi}_{rsf} + \tilde{\bm{\psi}} + \bm{\epsilon}}.
\end{equation}
Once $\tilde{\bm{\psi}}$ is a vector of the same length of $\bm{\epsilon}$, it is not possible to estimate both of them but just the sum. Let's call $\tilde{\bm{\psi}} + \bm{\epsilon}$ as $\bm{\epsilon}_{rsf}$ and finally our final model is given by Equation \eqref{eq:frailty_decomposition_final}.
\begin{equation} \label{eq:frailty_decomposition_final}
  h_{\theta}(t) = h_{\theta}^0(t)\exp\chav{\bm{X\beta}_{rsf} + \bm{\Psi}_{rsf} + \bm{\epsilon}_{rsf}}.
\end{equation}
Our main aim is to fit the restricted spatial frailty model. However, we would like to have estimates of the unrestricted model as well as the restricted model estimates. Therefore, we need to find equivalences between the restricted quantities and the unrestricted ones. With this equivalence, it is possible to have samples from both models concurrently. Equation~\eqref{eq:frailty_equivalence} presents this equivalence.
\begin{align} \label{eq:frailty_equivalence}
    h_{\bm{\theta}}(t) & = h_{\bm{\theta}}^0(t)\exp\chav{\bm{X\beta}_{rsf} + \bm{\Psi}_{rsf} + \bm{\epsilon}_{rsf}} \\
                  & = h_{\bm{\theta}}^0(t)\exp\chav{\bm{X\beta}_{rsf} + \bm{\Psi}_{rsf} + \tilde{\bm{\psi}} + \bm{\epsilon}_{sf}} \nonumber \\
                  & = h_{\bm{\theta}}^0(t)\exp\chav{\bm{X\beta}_{rsf} + \bm{P^{\bot}\Psi}_{sf} + \bm{\epsilon}_{sf}} \nonumber \\
                  & = h_{\bm{\theta}}^0(t)\exp\chav{\bm{X(\beta}_{rsf} - (\bm{X}^T\bm{X})^{-1}\bm{X}^T\bm{\Psi}_{sf}) + \bm{\Psi}_{sf} + \bm{\epsilon}_{sf}} \nonumber \\
                  & = h_{\bm{\theta}}^0(t)\exp\chav{\bm{X\beta}_{sf} + \bm{\Psi}_{sf} + \bm{\epsilon}_{sf}}, \nonumber
\end{align}
\noindent where ``sf'' means ``spatial frailty'' and represent the conventional spatial method and ``rsf'' means ``restricted spatial frailty'' and represents the model referred in Equation~\eqref{eq:frailty_decomposition_final}.

Given the unrestricted model, we can calculate the restricted quantities since $\bm{\beta}_{rsf} = \bm{\beta_{sf}} + (\bm{X}^T\bm{X})^{-1}X^T\bm{\Psi}_{sf}$, $\bm{P}^{\bot}\bm{\Psi}_{sf} = \bm{\Psi}_{rsf} + \tilde{\bm{\psi}}$ and $\bm{\epsilon}_{rsf} = \bm{\epsilon}_{sf} + \tilde{\bm{\psi}}$. With these equivalences it is possible to have estimates of all parameters of the restricted model. The general formulation in Equation~\eqref{eq:frailty_equivalence} shows how to obtain the restricted models estimates for the proportional hazard family including the Cox model (when $h_{\bm{\theta}}^0(t)$ is not defined). In other words, we just need a sample from the unrestricted model to get estimates from both unrestricted and restricted models. These results are applied for the entire family of proportional hazards models. It is important to notice that even if we fit a model without the $\bm{\epsilon}$ (independent) term, under the restricted model, the component $\bm{\epsilon}_{rsf}$ will appear.

\subsection{Reduction operator}

Although enlarging the spatial effect vector is a straightforward solution in Equation~\eqref{eq:decomposed_frailty_model}, the projection approach requires, for each element of a posterior sample, the calculations: 
\begin{itemize}
    \item $\bm{\beta}_{rsf} = \bm{\beta}_{sf} + (\bm{X}^T\bm{X})^{-1}\bm{X}^T\bm{\Psi}_{sf}$,
    \item $\bm{P}^{\bot}\bm{\Psi}_{sf} = (\bm{I}-\bm{X}(\bm{X}^T\bm{X})^{-1}\bm{X}^T)\bm{\Psi}_{sf} = \bm{\Psi}_{rsf} + \tilde{\bm{\psi}}$,
\end{itemize}
\noindent which requires products of matrices with support equal to the sample size ($N$). 

It is not unusual to work with data sets in which, for each area, several individuals are observed. As this number increases, the total sample size $N$ also increases, but not the number of areas, $n$, that remains fixed. Said that, the computation of the restricted model increases as $N$ increases. However, because $\bm{\Psi}_{sf}$ is constant by area, it is possible to get the same desired results but computing it with a reduced version of $\bm{P}$ and $\bm{P}^{\bot}$ matrices in which the new matrices are $(n\times n)$-dimensional instead of $(N\times N)$-dimensional. 

Let's define an operator that will help us to achieve the computational improvement. Let $\bm{X}_{N\times p}$ be a matrix with entries $X_{ijk}$ for an index $i$, an element $j$ and column $k$, and $\bm{G}_{N\times 1}$ is a vector of indices indicating, for each row of $\bm{X}_{N\times p}$, an index $i$ in a set of indices starting from $1$ until $n$ ($n \ll N$). Then the reduction operator $\circled{r}$ is defined by:
\begin{equation}
    \bm{X}_{N\times p} \circled{r} \bm{G} = \bm{x}_{n\times p},
\end{equation}
\noindent in which $\displaystyle x_{ik} = \sum_{j = 1}^{n_i}X_{ijk}$, and $n_i$ is the number of elements related with index $i$. This operator has several properties that allow us to simplify the computational procedure. Let $c$ be a constant, $\bm{r}_{n\times 1}$ is a column vector, $R = [r_{G_1}, \ldots, r_{G_N}]^T$ is a $N\times 1$ vector with repeated entries for each index of $\bm{G}$ (constant by indices), $\bm{P}_{p\times p}$ is a squared matrix and, $\bm{Q}_{m\times p}$ is a matrix. Therefore, the following properties are true:
\begin{enumerate}
    \item $(\bm{X}_1 + \bm{X}_2) \circled{r} \bm{G} = (\bm{X}_1 \circled{r} \bm{G}) + (\bm{X}_2 \circled{r} \bm{G})$,
    \item $(c\bm{X}) \circled{r} \bm{G} = c(\bm{X} \circled{r} \bm{G})$,
    \item $\bm{X^TR} = (\bm{X} \circled{r} \bm{G})^T\bm{r}$,
    \item $(\bm{QX^T}) \circled{r} \bm{G}^T = \bm{Q}(\bm{X} \circled{r} \bm{G})^T$,
    \item $(\bm{XPX^T}) \circled{r} \bm{G} = (\bm{X} \circled{r} G)\bm{P}\bm{X}^T$,
    \item $((\bm{XPX^T}) \circled{r} \bm{G}) \circled{r} \bm{G}^T = (\bm{X} \circled{r} \bm{G})\bm{P}(\bm{X} \circled{r} \bm{G})^T$,
    \item $(\bm{XPX^TR}) \circled{r} \bm{G} = (\bm{X} \circled{r} \bm{G})\bm{P}(\bm{X} \circled{r} \bm{G})^T\bm{r}$.
\end{enumerate}

The proofs of these properties are in Appendix \ref{apendix:proofs}. Using the reduction operator it is possible to compute $\bm{\beta_{rsf}}$ efficiently (by property 4):
\begin{align} \label{eq:beta_rsf}
\bm{\beta}_{rsf} & = \bm{\beta}_{sf} + (\bm{X^TX})^{-1}\bm{X^T\Psi}_{sf} \\
                 & = \bm{\beta}_{sf} + (\bm{X^TX})^{-1}(\bm{X} \circled{r} \bm{G})^T\bm{\psi}_{sf} \nonumber,
\end{align}
\noindent that is a product on a smaller dimension because $(\bm{X}^T\bm{X})^{-1}(\bm{X} \circled{r} \bm{G})^T$ is a $p\times n$ matrix. 

Also, to compute $\bm{\psi}_{rsf}$, using properties 1, 3 and 4, and defining $\bm{N}_{N\times N}$ as a diagonal matrix with $N_{ii} = n_{G_i}$ being the number of elements in each area, and $\bm{n}_{n\times n}$ being a diagonal matrix with $n_{jj} = n_{G_j}$ it is the same as 
\begin{align} \label{eq:psi_rsf}
\bm{\psi}_{rsf} & = \bm{N}^{-1}(\bm{I_N}-\bm{X}(\bm{X}^T\bm{X})^{-1}\bm{X}^T)\bm{\Psi}_{sf} \circled{r} \bm{G}\\
        & = (\bm{N}^{-1}\bm{\Psi}_{sf}-\bm{N}^{-1}\bm{X}(\bm{X}^T\bm{X})^{-1}\bm{X}^T\bm{\Psi}_{sf}) \circled{r} \bm{G} \nonumber \\
        & = (\bm{N}^{-1}\bm{\Psi}_{sf}) \circled{r} \bm{G} -\bm{N}^{-1}(\bm{X}(\bm{X}^T\bm{X})^{-1}\bm{X}^T\bm{\Psi}_{sf}) \circled{r} \bm{G} \nonumber \\
        & = (\bm{N}^{-1} \circled{r} \bm{G})^T\bm{\psi}_{sf} \circled{r} \bm{G} -\bm{N}^{-1}\bm{X}(\bm{X}^T\bm{X})^{-1}(\bm{X} \circled{r} \bm{G})^T\bm{\psi}_{sf} \circled{r} \bm{G} \nonumber \\
        & = \bm{\psi}_{sf} -(\bm{N}^{-1}\bm{X} \circled{r} \bm{G})(\bm{X}^T\bm{X})^{-1}(\bm{X} \circled{r} \bm{G})^T\bm{\psi}_{sf} \nonumber \\
        & = (\bm{I_n} - \bm{n}^{-1}(\bm{X} \circled{r} \bm{G})(\bm{X}^T\bm{X})^{-1}(\bm{X}^T\circled{r} \bm{G}))\bm{\psi}_{sf}. \nonumber
\end{align}
Then, for both $\bm{\beta}_{rsf}$ and $\bm{\psi}_{rsf}$ it is possible to calculate their values using small length matrices which is computationally attractive.

\section{Simulation} \label{sec:simulation}

This study is divided into two sections. First, the computational improvement will be presented by a simulation study that shows the reduction operator efficiency. Next, the capacity to recover the model parameters and the efficiency of the proposed correction will be evaluated. The methodology presented here does not depend on the method used to get samples from the unrestricted model. We provide one implementation for our methodology in the \texttt{RASCO} \texttt{R} package (\href{https://github.com/douglasmesquita/RASCO}{https://github.com/douglasmesquita/RASCO}). The package relies on the \texttt{R-INLA} package for inference because its computational benefits. It uses the \texttt{inla.posterior.sample} function to generate posterior samples of the parameters involved. This function allows us to have a sample from the approximated posterior distribution. The hyperparameters are sampled from the grid used in the numerical integration and the latent field is sampled from the Gaussian approximation conditioned on the hyperparameters. Although it is an approximation and not exact like a MCMC procedure, as will be shown in this section, the obtained results are reliable and takes advantage of the computation efficiency of the INLA method. Based on the posterior sample of the unrestricted model and using Equation~\eqref{eq:frailty_equivalence} we obtained posterior samples from the restricted model. Finally, the method does not depend on the parametric model chosen for the baseline hazard. Therefore, we opted to choose the widely applied Weibull proportional hazard model.

\subsection{Computational improvement} \label{subsec:comp_time}

To show the computational improvement using the reduction operator, we performed a simulation study. The time spent to get samples from the restricted model using the methodology described in Section~\ref{chap:survival} and the time spent applying the reduction operator were recorded.

The data were generated from the Weibull proportional hazard model for a spatial structure (polygons) containing $92$ areas. We vary the number of individuals in each area in the following grid: $2, 4, 8, 16, 32, 64 ~\text{and}~ 128$. Therefore, the total sample size $N$ is $92\times 2 = 184$ in the first scenario and $92\times 128 = 11,776$ in the last one. For each case, a posterior sample of size $5000$ was obtained and the restricted estimates were made based on it. This is a two-step technique, first, we get samples from the unrestricted model and then by Equation~\eqref{eq:frailty_equivalence} we get samples from the restricted model. Thus, we are able to record the time to fit the model and also the time to perform the correction. It is interesting to notice that, in both cases, the time spent to fit the unrestricted model should be the same and therefore we are not reporting it. Using the reduction operator, the correction step has always the same length matrix although the matrix $\bm{X}$ becomes larger at each step.

\begin{figure}[h]
    \centering
    \includegraphics[width = 0.95\textwidth]{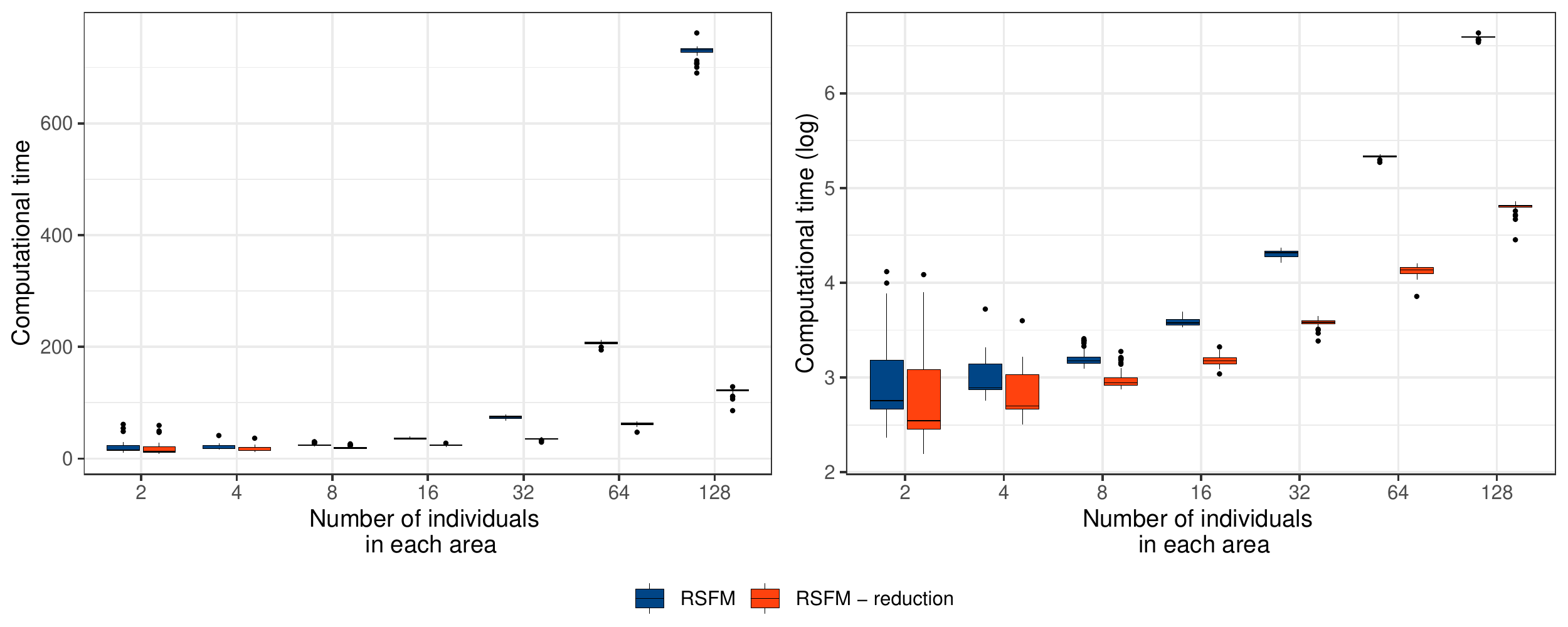}
    \caption{Time spent to fit the Weibull proportional hazard model with and without the reduction operator in seconds. Right: Original scale in seconds; Left: Logarithmic scale.}
    \label{fig:reduction_operator}
\end{figure}

Figure~\ref{fig:reduction_operator} shows the computational cost for applying these two approaches, varying according to the number of subjects in each area. The boxplots are based on $100$ repetitions. 
As one can see, the computational cost is increasing as $N$ increases for the model without the reduction step. The increment in time for the pure model increases drastically because for each posterior sample we must calculate $\bm{P^{\bot}}\bm{\Psi}_{sf}$. This is a product of a $N\times N$ matrix by a $N \times 1$ vector (this product is repeated $5,000$ times). Instead, the model with the reduction operator calculates $((\bm{P}^{\bot} \circled{r} \bm{G}) \circled{r} \bm{G})\bm{\psi}_{sf}$ which is a product of a $n\times n$ matrix by a $n \times 1$ vector). 

The time spent to calculate $(\bm{P}^{\bot} \circled{r} \bm{G}) \circled{r} \bm{G}$ also increases as $N$ increases, but this calculation occurs just once. Also, it is a straightforward calculation that is not strongly affected by the sample size. Thus, since the computational cost to calculate the reduced model is preferable and the fact that the results are strictly the same, we will use the reduction operator for the rest of the work.

\subsection{Confounding alleviation} \label{subsec:conf_correction}

To evaluate the model ability to estimate the parameters, the data were generated from the Weibull proportional hazard model
\begin{align}
    & h(t_i) =\alpha t_i^{\alpha - 1}\exp\chav{\beta_1X_{i1} + \beta_2X_{i2} + \psi_i}, \\
    & \bm{\psi} \sim \text{ICAR}(\bm{W}, \tau_{\psi}), \nonumber \nonumber
\end{align}
\noindent where $\alpha = 1.2$, $\beta_1 = -0.3$, $\beta_2 = 0.3$ and $\tau_{\psi} = 0.75$. To evaluate the performance in terms of recovering the parameters in this model, we are considering the right censoring scheme and 4 censoring levels: 0\%, 25\%, 50\% and 75\%.

We generated 1,000 datasets under each setup and 2 scenarios: 1) $\bm{X}_1$ and $\bm{X}_2$ are random variables and therefore no spatial confounding is expected; 2) $\bm{X}_1$ is a random variable but $\bm{X}_2$ is the set of centroids' latitudes of each county. The set of weakly informative priors was taken as follow

$\alpha \sim \Gamma(0.001, 0.001)$, 

$\beta_{j} \sim \text{Normal}(0, 0.001), ~ j = 0, 1, 2,$

$\tau_{\psi} \sim \Gamma(0.5, 0.0005).$

Table \ref{tab:simulation_frailty} presents the mean of the estimated values (Mean), the mean of the standard deviations (SD), the coverage rate for a nominal rate of 95\%  (Cov) and the mean squared error (MSE) for each scenario.
\begin{table}[h]
\caption{Simulation results the spatial frailty model experiment. The results are shown by mean, standard deviation (SD), coverage rate for a nominal rate of 95 \% (Cov) and mean square error (MSE).}
\label{tab:simulation_frailty}
\centering
\resizebox{.99\textwidth}{!}{%
\begin{tabular}{@{}llllllllllllll@{}}
\toprule
\multirow{3}{*}{Censure} & \multirow{3}{*}{Parameter} & \multicolumn{6}{l}{Without spatial confounding}                   & \multicolumn{6}{l}{with spatial confounding}                     \\ \cmidrule(l){3-14} 
                         &                            & SFM          &         &        & RSFM         &         &        & SFM          &         &        & RSFM         &         &        \\ \cmidrule(l){3-14} 
                         &                            & Mean (SD)    & COV     & MSE    & Mean (SD)    & COV     & MSE    & Mean (SD)    & COV     & MSE    & Mean (SD)    & COV     & MSE    \\ \midrule
00.00\%                  & $\alpha$                   & 1.18 (0.07)  & 83.20\% & 0.0071 & 1.18 (0.07)  & 83.20\% & 0.0071 & 1.17 (0.07)  & 77.80\% & 0.0087 & 1.17 (0.07)  & 77.80\% & 0.0087 \\
                         & $\beta_1$                  & 0.29 (0.08)  & 93.90\% & 0.0060 & 0.29 (0.07)  & 94.80\% & 0.0053 & 0.29 (0.08)  & 93.60\% & 0.0061 & 0.29 (0.07)  & 94.90\% & 0.0053 \\
                         & $\beta_2$                  & -0.29 (0.08) & 93.70\% & 0.0062 & -0.30 (0.07) & 93.80\% & 0.0054 & -0.28 (0.20) & 83.40\% & 0.0609 & -0.29 (0.07) & 93.00\% & 0.0062 \\ \midrule
25.00\%                  & $\alpha$                   & 1.20 (0.08)  & 80.40\% & 0.0097 & 1.20 (0.08)  & 80.40\% & 0.0097 & 1.18 (0.08)  & 69.10\% & 0.0118 & 1.18 (0.08)  & 69.10\% & 0.0118 \\
                         & $\beta_1$                  & 0.30 (0.09)  & 93.70\% & 0.0082 & 0.30 (0.08)  & 94.40\% & 0.0073 & 0.29 (0.09)  & 92.80\% & 0.0083 & 0.29 (0.08)  & 94.40\% & 0.0072 \\
                         & $\beta_2$                  & -0.30 (0.09) & 92.80\% & 0.0083 & -0.30 (0.08) & 93.60\% & 0.0074 & -0.28 (0.19) & 77.10\% & 0.0723 & -0.29 (0.09) & 93.60\% & 0.0085 \\ \midrule
50.00\%                  & $\alpha$                   & 1.21 (0.09)  & 80.00\% & 0.0137 & 1.21 (0.09)  & 80.00\% & 0.0137 & 1.17 (0.08)  & 72.80\% & 0.0131 & 1.17 (0.08)  & 72.80\% & 0.0131 \\
                         & $\beta_1$                  & 0.30 (0.10)  & 94.30\% & 0.0116 & 0.30 (0.10)  & 93.60\% & 0.0106 & 0.29 (0.10)  & 93.60\% & 0.0114 & 0.29 (0.10)  & 95.10\% & 0.0104 \\
                         & $\beta_2$                  & -0.30 (0.11) & 93.20\% & 0.0129 & -0.30 (0.10) & 94.10\% & 0.0112 & -0.28 (0.17) & 63.80\% & 0.0936 & -0.28 (0.11) & 93.10\% & 0.0131 \\ \midrule
75.00\%                  & $\alpha$                   & 1.20 (0.11)  & 82.30\% & 0.0185 & 1.20 (0.11)  & 82.30\% & 0.0185 & 1.18 (0.11)  & 83.30\% & 0.0158 & 1.18 (0.11)  & 83.30\% & 0.0158 \\
                         & $\beta_1$                  & 0.30 (0.14)  & 92.20\% & 0.0223 & 0.30 (0.14)  & 93.30\% & 0.0211 & 0.30 (0.14)  & 93.20\% & 0.0210 & 0.30 (0.14)  & 94.30\% & 0.0194 \\
                         & $\beta_2$                  & -0.29 (0.14) & 93.10\% & 0.0236 & -0.29 (0.14) & 93.80\% & 0.0215 & -0.30 (0.17) & 56.30\% & 0.1418 & -0.30 (0.15) & 92.80\% & 0.0264 \\ \bottomrule
\end{tabular}
}
\end{table}
It is possible to observe that, without spatial confounding, the SFM (Spatial Frailty Model) and the RSFM (Restricted Spatial Frailty Model) approaches present similar values for mean, coverage and mean squared error. Under spatial confounding, it is possible to observe that the point estimates (mean) are accurate for all parameters and both models. However, the standard deviation of $\beta_2$ is, on average, greater for the SFM model than for the RSFM model. Also, it is possible to observe that the MSE of $\beta_2$ is greater for the SFM than for the RSFM. The coverage rate seems adequate for both models except for the parameter $\alpha$. However, this inconsistency is observed in all models and for the INLA method $\alpha$ is considered as a hyperparameter with marginal posterior approximation obtained by numerical integration. This can be a possible explanation for why the \texttt{R-INLA} package might not be providing an appropriate coverage for this parameter. Anyhow, the true understanding of the reason that is causing this phenomenon is out of the scope of the current manuscript and does not affect our analysis and conclusions.

a pursue of the investigation of the reason that is causing this phenomenon is out of the scope of the current manuscript.

For the INLA method $\alpha$ is a hyperparameter and its posterior approximation is obtained by numerical integration that might now be providing an appropriate approximation in this case. However,  this inconsistency is observed in all models a pursue of the investigation of the reason that is causing this phenomenon is out of the scope of the current manuscript.

Another interesting conclusion is that as much the censure level increases the standard deviations also increase in all cases. It is showing that, in those models with bigger censoring rates, the estimates are less accurate as expected.

\begin{figure}[h]
    \centering
    \includegraphics[width = 0.99\textwidth]{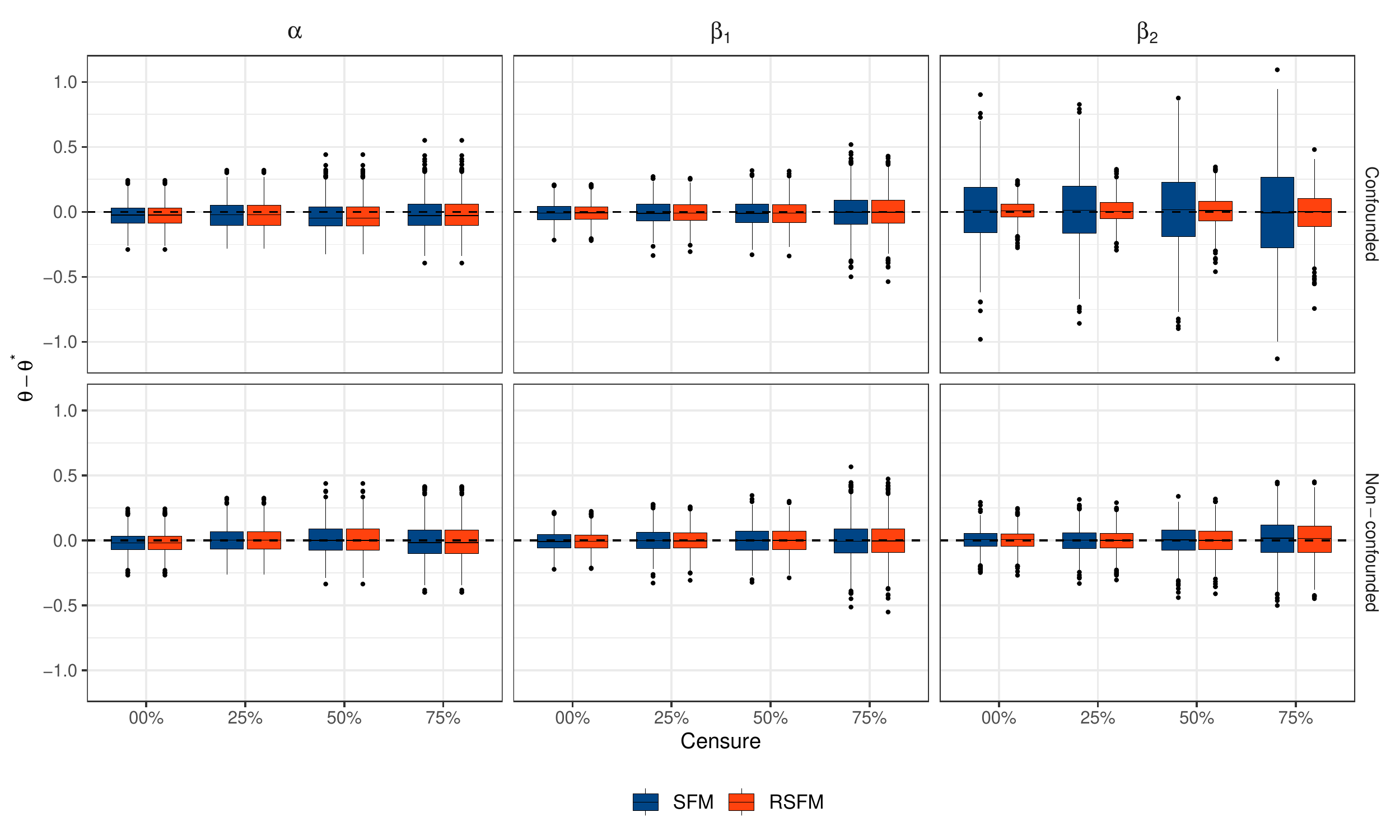}
    \caption{Boxplot of $(\bm{\theta} - \bm{\theta}^{\ast})$ for $\bm{\theta} = \{\alpha, \beta_1, \beta_2 \}$ in the spatial frailty model. Dashed line represents the value 0.}
    \label{fig:parameters_frailty}
\end{figure}

The projection-based approach aims to estimate $\bm{\beta}^{\ast} = \bm{\beta}_{sf} + \bm{P}^{\bot}\bm{\Psi}$. In this case, we reported Figure \ref{fig:parameters_frailty} for $\displaystyle (\bm{\theta} - \bm{\theta}^{\ast})$ where $\bm{\theta} = \{\alpha, \beta_1, \beta_2\}$.
We expect all the values to be around 0, which means that the estimate is not biased. We can see, for $\alpha$ and $\beta_1$, that the estimates are similar for both SFM and RSFM models. However, for $\beta_2$ the behavior changes for the model with and without confounding. In the model without spatial confounding, as expected, $\beta_2$ behaves in the same way for $\beta_1$. For the model with spatial confounding, we can observe that the $\displaystyle (\bm{\theta} - \bm{\theta}^{\ast})$ tends to be around 0 for the SFM and also for the RSFM. Although they are centered at 0, the dispersion of SFM seems to be bigger than the dispersion of the RSFM model which, in this case, suggests variance inflation.

From the perspective of the level of censorship, we can see a smooth increment in the coefficients' variance for all cases. This result is explained by the fact that, with the increment of the censoring rate, we have less cases and therefore less information about the effect of each covariate.  

\begin{figure}[h]
    \centering
    \includegraphics[width = 0.99\textwidth]{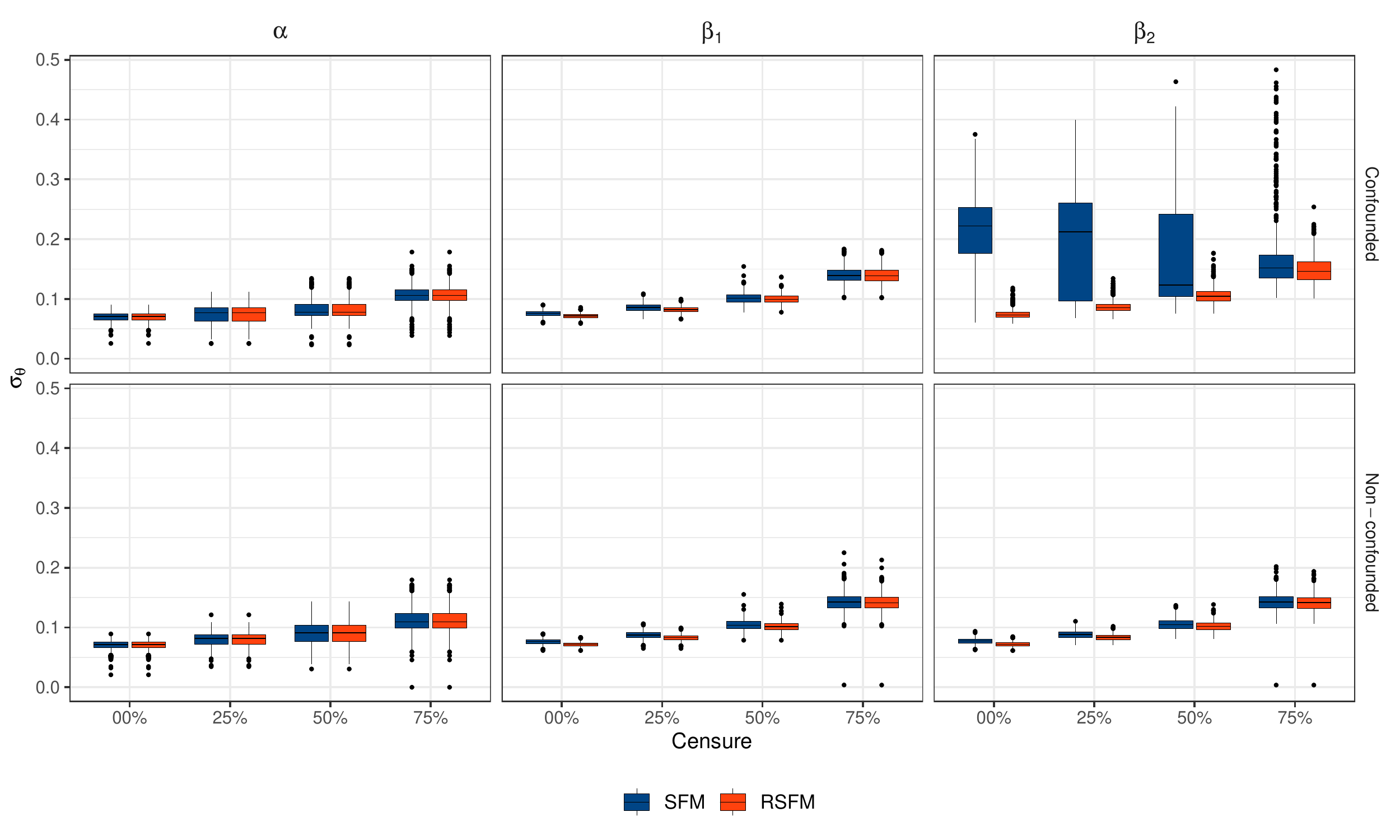}
    \caption{Boxplot of $\displaystyle \sigma_{\theta}$ for $\bm{\theta} = \{\beta_{1}, \beta_{2}\}$ for the SFM and RSFM where $\sigma_{\theta}$ represents the standard deviation of $\theta$.}
    \label{fig:parameters_frailty_sd}
\end{figure}

\begin{figure}[h]
    \centering
    \includegraphics[width = 0.99\textwidth]{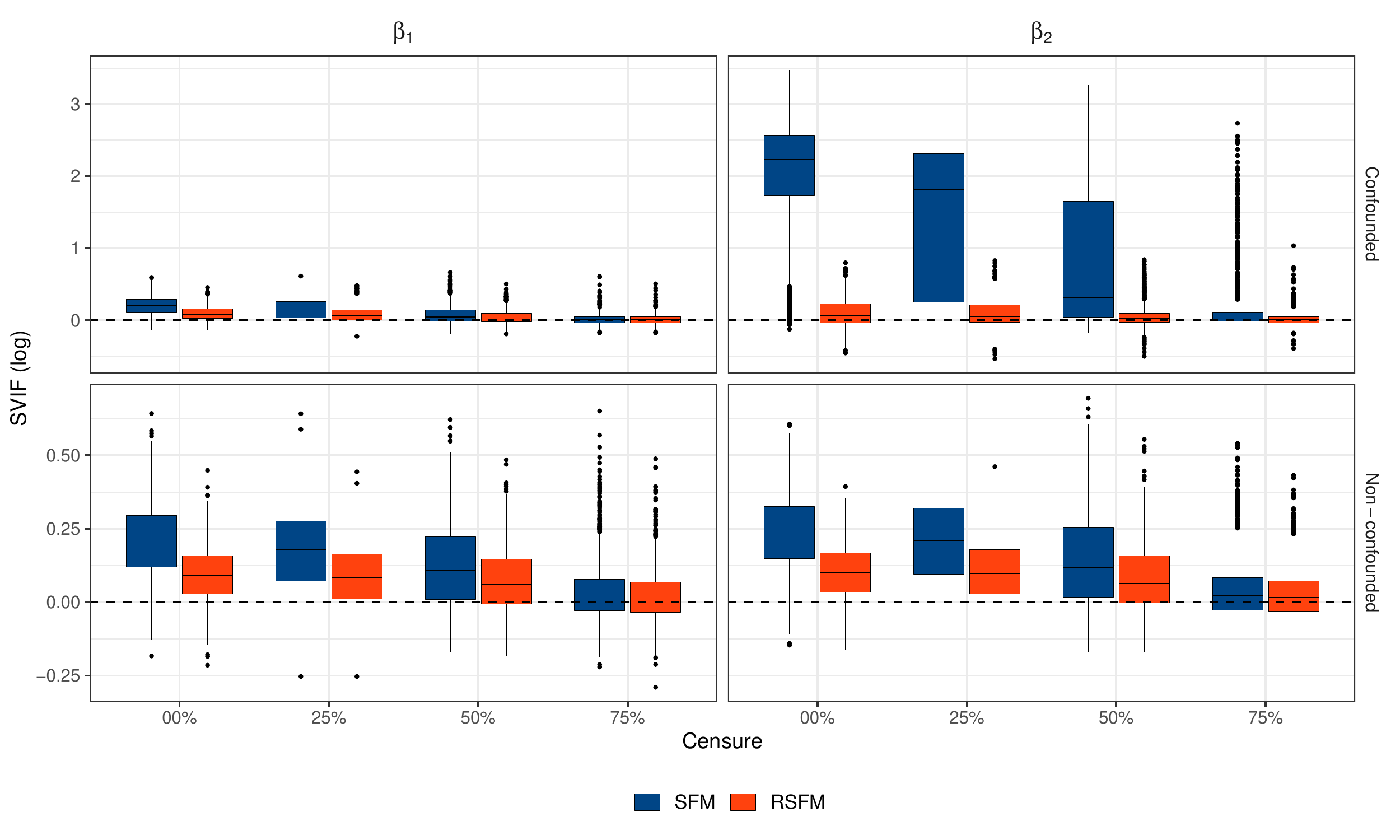}
    \caption{Boxplot of the SVIF (log scale) between spatial models (SFM and RSFM) and the baseline model (Weibull proportional hazard model). Dashed line marks the value 0, which in the log scale represent the equality of variances.}
    \label{fig:parameters_frailty_vif}
\end{figure}

Figures \ref{fig:parameters_frailty_sd}, \ref{fig:parameters_frailty_vif}, show for SFM and RSFM, the standard deviations and the SVIF (defined in Section \ref{subsec:svif}) comparing with the non-spatial model. An SVIF equal to 1 indicates that the variances of both models are the same. However, because the spatial model is more complex it is expected an increment in the variance.
We can observe in Figure \ref{fig:parameters_frailty_sd} that the standard deviations are similar in all cases except for $\beta_2$ in the scenario with spatial confounding. In this case, we can see that the higher the level of censorship, the more similar is the standard deviation. 
In Figure~\ref{fig:parameters_frailty_vif}, we can observe a downward trend in the SVIF for both models when the level of censorship increases. It means that the efficiency of the correction decreases with the increment of the censured individuals, which is in agreement with Figure \ref{fig:parameters_frailty}.

The behavior for $\beta_1$ and $\beta_2$ in the model without spatial confounding are similar and this is also true for the behavior of $\beta_1$ under spatial confounding. However, the parameter $\beta_2$ under spatial confounding presents huge inflation of variances for the model without correction. In some cases, we experienced a variance $\exp\{3\} \approx 20$ times bigger. In these cases, the restricted model behaves well and it keeps the variance stable.

\section{Respiratory cancer in California} \label{sec:application}

To fit the model, we use a right censoring scheme with the Weibull proportional hazard model. Our baseline model is the Non-spatial (NS) model given by the Weibull proportional hazard model and five covariates: 1) gender; 2) race; 3) disease stage; 4) age at diagnosis; 5) the percentage of people who smoke every day or most days (areal level). The spatial frailty model (SFM) also includes the ICAR spatial term, and the restricted spatial frailty model (RSFM) alleviates possible spatial confounding. 

In Table \ref{tab:frailty_results}, $\alpha$ is the shape parameter of the Weibull distribution and the estimate was almost the same in the NS and SFM models (RSFM estimate is the same of the SFM for hyperparameters). The parameter $\tau_w$ represents the precision for the ICAR model. The other parameters are related with the covariates in the modeling.

\begin{table}[]
\caption{Time until death by lung and bronchus cancer in California (US). Results are presented as mean, standard deviation (SD) and the 95 \%credibility interval (ICr).}
\label{tab:frailty_results}
\centering
\resizebox{.95\textwidth}{!}{
\begin{tabular}{@{}lllllll@{}}
\toprule
\multirow{2}{*}{Parameter} & NS             &                & SFM           &                & RSFM            &                \\ \cmidrule(l){2-7} 
                           & Mean (SD)      & ICr            & Mean (SD)      & ICr            & Mean (SD)      & ICr            \\ \midrule
$\alpha$                   & 0.85 (0.0032)  & (0.85; 0.86)   & 0.86 (0.0033)  & (0.85; 0.86)   & 0.86 (0.0033)  & (0.85; 0.86)   \\
$\tau_{\psi}$              &                &                & 22.75 (6.8306) & (10.99; 36.49) & 22.75 (6.8306) & (10.99; 36.49) \\
$\beta_0$                  & -4.07 (0.2048) & (-4.44; -3.65) & -3.58 (0.2091) & (-4.02; -3.20) & -4.06 (0.2001) & (-4.47; -3.69) \\
Gender                     &                &                &                &                &                &                \\
~~~~ Female                     & ref.           &                & ref.           &                & ref.           &                \\
~~~~ Male                       & 0.19 (0.0099)  & (0.17; 0.21)   & 0.19 (0.0098)  & (0.18; 0.21)   & 0.19 (0.0098)  & (0.18; 0.21)   \\
Race                       &                &                &                &                &                &                \\
~~~~ Non-black                  & ref.           &                & ref.           &                & ref.           &                \\
~~~~ Black                      & 0.16 (0.0177)  & (0.13; 0.20)   & 0.17 (0.0176)  & (0.13; 0.20)   & 0.17 (0.0175)  & (0.13; 0.20)   \\
Cancer stage               &                &                &                &                &                &                \\
~~~~ In situ                    &                &                &                &                &                &                \\
~~~~ Localized                  & 1.51 (0.1985)  & (1.12; 1.87)   & 1.51 (0.1940)  & (1.13; 1.90)   & 1.51 (0.1940)  & (1.13; 1.90)   \\
~~~~ Regional                   & 2.60 (0.1984)  & (2.20; 2.96)   & 2.60 (0.1935)  & (2.25; 3.01)   & 2.60 (0.1935)  & (2.25; 3.02)   \\
~~~~ Distant                    & 3.73 (0.1985)  & (3.35; 4.11)   & 3.74 (0.1936)  & (3.37; 4.14)   & 3.74 (0.1936)  & (3.37; 4.14)   \\
Age at diagnosis           & 0.02 (0.0005)  & (0.02; 0.02)   & 0.02 (0.0005)  & (0.02; 0.02)   & 0.02 (0.0005)  & (0.02; 0.02)   \\
\% Smokers           & 2.13 (0.1821)  & (1.79; 2.49)   & -0.90 (0.3848) & (-1.68; -0.21) & 2.13 (0.1819)  & (1.79; 2.51)   \\ \bottomrule
\end{tabular}
}
\end{table}
From the epidemiological point of view, the NS model reflects the theory that patients in a more advanced stage of the disease have a higher risk of death (In situ $<$ Localized $<$ Regional $<$ Distant). Also, males have a higher risk when compared to females. Same way, black people have a higher risk when compared with non-black people. Further, the older the individual the greater is the risk. The coefficient for the percentage of smokers in the county indicates an increment in the risk of death due to lung and bronchus cancer. This covariate is our best proxy about individual tobacco consumption.

When we compare the results from the NS model with those of the SFM, one can notice that for gender, race, stage of the disease and age at diagnosis, the results are similar with small differences in the estimates. However, for the coefficient of the percentage of smokers, the point estimate changes drastically and there is variance inflation (variance is about 5 times greater for the SFM). In the SFM model the credibility interval changes drastically pointing that the percentage of smokers is a protective factor for lung and bronchus cancer death. The restricted spatial frailty model (RSFM) was applied and we can notice that it returns similar estimates to those from the NS model, as expected. The credibility interval is now pointing that the higher the percentage of smokers, higher is the risk for cancer death.   

\begin{figure}[h]
    \centering
    \includegraphics[width = 0.95\textwidth]{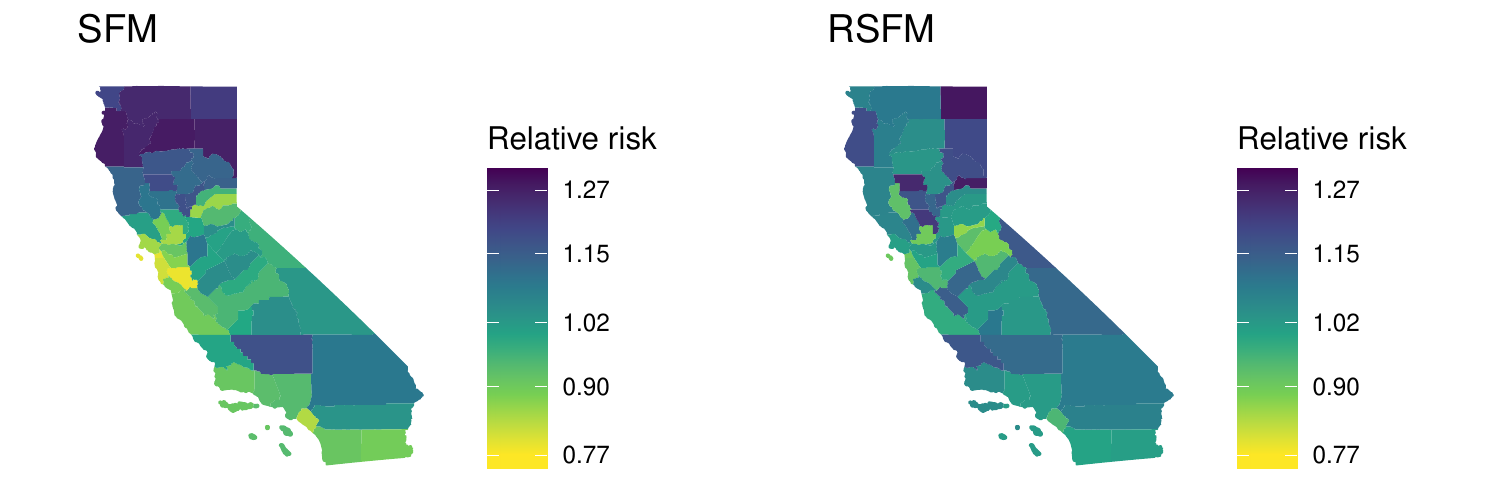}
    \caption{Spatial risk effects for death by lung and bronchus cancer in California (US).}
    \label{fig:frailty_effects}
\end{figure}

Figure~\ref{fig:frailty_effects} shows the spatial effect $\exp\{\bm{\psi}\}$ for the SFM and RSFM.
We can see that the patterns are smoother for the RSFM case. However, the pattern remains similar to the SFM model being higher in the north of the state, less intense in the center and again high in the south. This result might be useful to create new policies or new health care centers for lung and bronchus cancer in California. 

We can conclude that the employment of the proposed restricted model is important in several ways. The first advantage is that the model conclusions retains the interpretability of the baseline model, keeping important conclusions about the model's covariates. Secondly, the computational improvement provided by the reduction operator appears as an important feature because it allows for the model to scale for large $N$. Third, it allows the user to apply its preferable software to get posterior samples from the unrestricted model and with its posterior sample correct for possible spatial confounding without refitting the model. Fourth, under spatial confounding, the variances of coefficients are not inflated as it is for the conventional model. Finally, the spatial pattern is similar when compared with the unrestricted model which shows that the correction maintain the underlying the spatial patterns.

\section{Final remarks} \label{sec:final_remarks}

Spatial confounding is a limitation of spatial models that needs attention since it can imply in wrong conclusions about important covariates effects. The conventional solution based on projections cannot be directly applicable for the spatial frailty models due to the fact that the support of fixed and random effects does not match. 

This work showed an alternative to alleviate the effects of spatial confounding in the spatial frailty framework. We provided an efficient way to fit the restricted spatial frailty model based on a posterior sample of the unrestricted model. To solve the difference in the supports of fixed and spatial effects, we proposed a reduction operator that is not only adequate to alleviate the spatial confounding but also has computational benefits. The method adequacy and efficiency were shown by a simulation study that proved its relevance and importance. 

We have applied the methodology in the data provided by the SEER. We enrich the data set with some county-level information provided by the CHRR. The spatial frailty model was employed to model the time until death by lung and bronchus cancer in California between 2010 and 2016. Our method provided alleviation of the spatial confounding while keeping the model interpretability. 

For future work, we may investigate the effects of spatio-temporal confounding in frailty models. To the best of our knowledge, up to now, no attention is paid to these extension. Moreover, the reduction operator seems an easy and applicable tool for statistical models. It can be directly employed for discrete models in which the math involves products of a matrix and a variable that is constant by groups (counts for example). Therefore, it is possible to think in a discretization of continuous variables aiming to reduce the computational effort keeping the desired accuracy. 

Finally, the proposed methodology is available in the \texttt{RASCO} \texttt{R} package with more methods and examples. Up to now the package is able to work with GLMM, Shared Component models (with two outcomes) and the aforementioned restricted spatial frailty model. The first version is available at \href{https://github.com/douglasmesquita/RASCO}{https://github.com/douglasmesquita/RASCO}. 

\section{Appendix} \label{apendix:proofs}

In this section we present the proofs of the properties of the reduction operator. Let $\bm{X}_{N\times p}$ be a matrix with entries $X_{ijk}$ for an index $i$, an element $j$ and column $k$, and $\bm{G}_{N\times 1}$ a vector of indices indicating for each line of $\bm{X}_{N\times p}$ an index $i$ in a set of indices starting from $1$ until $n$ ($n \ll N$). Then the reduction operator $\circled{r}$ is defined by:
\begin{equation}
    \bm{X}_{N\times p} \circled{r} \bm{G} = \bm{x}_{n\times p},
\end{equation}
in which $\displaystyle x_{ik} = \sum_{j = 1}^{n_i}X_{ijk}$, and $n_i$ is the number of elements associated with index $i$.

For $\bm{X}_1$ and $\bm{X}_2$, $N\times p$ matrices with entries $X_{dijk}$ for $d = 1, 2$, it is true that $(\bm{X}_1 + \bm{X}_2) \circled{r} \bm{G} = (\bm{X}_1 \circled{r} \bm{G}) + (\bm{X}_2 \circled{r} \bm{G})$.
\begin{proof}
Consider the general term

\begin{align*}
\PR{(\bm{X}_1 + \bm{X}_2) \circled{r} \bm{G}}_{ik} = & \sum_{j = 1}^{n_i}\PC{X_{1ijk} + X_{2ijk}} \\
                                                   = & \sum_{j = 1}^{n_i}X_{1ijk} + \sum_{j = 1}^{n_i}X_{2ijk} \\
                                                   = & \PR{\bm{X}_1 \circled{r} \bm{G}}_{ik} + \PR{\bm{X}_2 \circled{r} \bm{G}}_{ik} \\
                                                   = & \PR{\PC{\bm{X}_1 \circled{r} \bm{G}} + \PC{\bm{X}_2 \circled{r} \bm{G}}}_{ik},
\end{align*}
$\forall i \in \chav{1, \ldots, n}, \forall k \in \chav{1, \ldots, p}$

$\implies (\bm{X}_1 + \bm{X}_2) \circled{r} \bm{G} = \PC{\bm{X}_1 \circled{r} \bm{G}} + \PC{\bm{X}_2 \circled{r} \bm{G}}$.
\end{proof}

For $\bm{X}_{N\times p}$ matrix with entries $X_{ijk}$ and a constant $c$, it is true that $(c\bm{X}) \circled{r} \bm{G} = c(\bm{X} \circled{r} \bm{G})$.
\begin{proof}
Consider the general term

\begin{align*}
\PR{(c\bm{X}) \circled{r} \bm{G}}_{ik} = & \sum_{j = 1}^{n_i}\PC{cX_{ijk}} \\
                                                   = & c\sum_{j = 1}^{n_i}X_{ijk} \\
                                                   = & c\PR{\bm{X} \circled{r} \bm{G}}_{ik} \\
                                                   = & \PR{c\PC{\bm{X} \circled{r} \bm{G}}}_{ik},
\end{align*}
$\forall i \in \chav{1, \ldots, n}, \forall k \in \chav{1, \ldots, p}$

$\implies (c\bm{X}) \circled{r} \bm{G} = c(\bm{X} \circled{r} \bm{G})$.
\end{proof}

For $\bm{X}_{N\times p}$ matrix with entries $X_{ijk}$, $\bm{r}_{n\times 1}$ a column vector, $R = [r_{G_1}, \ldots, r_{G_N}]^T$ a $N\times 1$ vector with repeated entries for each index of $\bm{G}$ (constant by indices), it is true that $\bm{X^TR} = (\bm{X} \circled{r} \bm{G})^T\bm{r}$.
\begin{proof}
Consider the general term

\begin{align*}
\PR{\bm{X}^T\bm{R}}_{k} = & \sum_{l = 1}^{N}\PC{X_{lk}R_l} \\
                  = & \sum_{i = 1}^{n}\sum_{j = 1}^{n_i}\PC{X_{ijk}r_i} \\
                  = & \sum_{i = 1}^{n}r_i\sum_{j = 1}^{n_i}X_{ijk} \\                                = & \sum_{i = 1}^{n}r_i\PR{\bm{X} \circled{r} \bm{G}}_{ik} \\
                  = & \PR{\bm{X} \circled{r} \bm{G}}_{1k}r_1 + \ldots + \PR{\bm{X} \circled{r} \bm{G}}_{nk}r_n \\
                  = & \PR{\PC{\bm{X} \circled{r} \bm{G}}^T}_{.k}\bm{r} \\
                  = & \PR{\PC{\bm{X} \circled{r} \bm{G}}^T\bm{r}}_k,
\end{align*}
$\forall k \in \chav{1, \ldots, p}$

$\implies \bm{X^TR} = (\bm{X} \circled{r} \bm{G})^T\bm{r}$.
\end{proof}

For $\bm{X}_{N\times p}$ matrix with entries $X_{ijk}$ and $\bm{Q}_{M\times p}$ a matrix with entries $Q_{mk}$, it is true that $(\bm{QX^T}) \circled{r} \bm{G}^T = \bm{Q}(\bm{X} \circled{r} \bm{G})^T$.
\begin{proof}
Consider the general term $\PR{\bm{Q}\bm{X}^T}_{ml} = \sum_{k = 1}^{p}Q_{mk}X_{lk}$, and

\begin{align*}
\PR{(\bm{Q}\bm{X}^T) \circled{r} \bm{G}^T}_{mi} = & \sum_{j = 1}^{n_i}\sum_{k = 1}^{p} Q_{mk} X_{ijk} \\
                                             = & \sum_{k = 1}^{p} Q_{mk}\sum_{j = 1}^{n_i} X_{ijk} \\
                                             = & \sum_{k = 1}^{p} Q_{mk}\PR{\PC{\bm{X} \circled{r} \bm{G}}}_{ik} \\
                                             = & Q_{m1}\PR{\PC{\bm{X} \circled{r} \bm{G}}}_{i1} + \ldots + Q_{mp}\PR{\PC{\bm{X} \circled{r} \bm{G}}}_{ip} \\
                                             = & \bm{Q}_{m.}\PR{\PC{\bm{X} \circled{r} \bm{G}}^T}_{.i} \\
                                             = & \PR{\bm{Q}\PC{\bm{X} \circled{r} \bm{G}}^T}_{mi},
\end{align*}
$\forall m \in \chav{1, \ldots, M}, \forall k \in \chav{1, \ldots, p}$

$\implies (\bm{QX^T}) \circled{r} \bm{G}^T = \bm{Q}(\bm{X} \circled{r} \bm{G})^T$.
\end{proof}

For $\bm{X}_{N\times p}$ matrix with entries $X_{ijk}$ and $\bm{P}_{p\times p}$ a squared matrix, it is true that $(\bm{XPX^T}) \circled{r} \bm{G} = (\bm{X} \circled{r} G)\bm{P}\bm{X}^T$.
\begin{proof}
\begin{align*}
\PC{\bm{X}\bm{P}\bm{X}^T} \circled{r} \bm{G} = & \PC{\bm{X}\bm{K}} \circled{r} \bm{G} \\
                                            = & \PC{\PC{\bm{K}^T\bm{X}^T} \circled{r} \bm{G}^T}^T \\
                   \stackrel{\text{1.4}}{=} & \PC{\bm{K}^T\PC{\bm{X} \circled{r} \bm{G}}^T}^T \\
                                            = & \PC{\bm{X} \circled{r} \bm{G}}\bm{K} \\
                                            = & \PC{\bm{X} \circled{r} \bm{G}}\bm{P}\bm{X}^T.
\end{align*}
\end{proof}

For $\bm{X}_{N\times p}$ matrix with entries $X_{ijk}$ and $\bm{P}_{p\times p}$ a squared matrix, it is true that  $((\bm{XPX^T}) \circled{r} \bm{G}) \circled{r} \bm{G}^T = (\bm{X} \circled{r} \bm{G})\bm{P}(\bm{X} \circled{r} \bm{G})^T$.
\begin{proof}
\begin{align*}
\PC{\PC{\bm{X}\bm{P}\bm{X}^T} \circled{r} \bm{G}} \circled{r} \bm{G}^T \stackrel{\text{1.5}}{=} & \PC{\PC{\bm{X} \circled{r} \bm{G}}\bm{P}\bm{X}^T} \circled{r} \bm{G}^T \\
                       = & \PC{\bm{K}\bm{X}^T} \circled{r} \bm{G}^T \\
\stackrel{\text{1.4}}{=} & \bm{K}\PC{\bm{X} \circled{r} \bm{G}}^T \\
                       = & \PC{\bm{X}\circled{r}\bm{G}}\bm{P}\PC{\bm{X} \circled{r} \bm{G}}^T.
\end{align*}
\end{proof}

For $\bm{X}_{N\times p}$ matrix with entries $X_{ijk}$, $\bm{r}_{n\times 1}$ a column vector, $R = [r_{G_1}, \ldots, r_{G_N}]^T$ a $N\times 1$ vector with repeated entries for each index of $\bm{G}$ (constant by indices) and $\bm{P}_{p\times p}$ a squared matrix, it is true that $(\bm{XPX^TR}) \circled{r} \bm{G} = (\bm{X} \circled{r} \bm{G})\bm{P}(\bm{X} \circled{r} \bm{G})^T\bm{r}$.
\begin{proof}
\begin{align*}
\PC{\bm{X}\bm{P}\bm{X}^T\bm{R}} \circled{r} \bm{G} = & \PC{\bm{X}\bm{K}} \circled{r} \bm{G} \\
                                                  = & \PC{\PC{\bm{K}^T\bm{X}^T} \circled{r} \bm{G}^T}^T \\
                           \stackrel{\text{1.4}}{=} & \PC{\bm{K}^T\PC{\bm{X} \circled{r} \bm{G}}^T}^T \\
                                                  = & \PC{\bm{X} \circled{r} \bm{G}}\bm{K} \\
                                                  = & \PC{\bm{X} \circled{r} \bm{G}}\bm{P}\bm{X}^T\bm{R} \\
                           \stackrel{\text{1.3}}{=} & \PC{\bm{X} \circled{r} \bm{G}}\bm{P}\PC{\bm{X} \circled{r} \bm{G}}^T\bm{r}.
\end{align*}
\end{proof}

\bibliographystyle{plainnat}
\bibliography{bibliografia} 


\end{document}